\documentclass[aps,prb,twocolumn,superscriptaddress]{revtex4-1}

\usepackage{graphicx}
\usepackage{ulem}
\usepackage{color}
\usepackage{multirow}
\usepackage{xspace}

\usepackage{amsmath}
\usepackage{siunitx}
\usepackage{miller}

\usepackage[version=3]{mhchem}
\usepackage{natbib}

\newcommand{\cbs}{Cu$_x$Bi$_2$Se$_{3}$}

\begin{document}


\title{Crystal structure and distortion of superconducting Cu$_x$Bi$_2$Se$_{3}$}

\author{Tobias Fr\"ohlich}
\affiliation{$I\hspace{-.1em}I$. Physikalisches Institut,
Universit\"at zu K\"oln, Z\"ulpicher Str. 77, D-50937 K\"oln,
Germany}

\author{Zhiwei Wang}
\affiliation{$I\hspace{-.1em}I$. Physikalisches Institut,
Universit\"at zu K\"oln, Z\"ulpicher Str. 77, D-50937 K\"oln,
Germany}
\affiliation{Key Laboratory of Advanced
Optoelectronic Quantum Architecture and Measurement (MOE),
School of Physics, Beijing Institute of Technology, Beijing, 10086, P. R. China}

\author{Mahasweta Bagchi}
\affiliation{$I\hspace{-.1em}I$. Physikalisches Institut,
Universit\"at zu K\"oln, Z\"ulpicher Str. 77, D-50937 K\"oln,
Germany}

\author{Anne Stunault}
\affiliation{Institut Laue Langevin, 6 Rue Jules Horowitz BP 156, F-38042 Grenoble CEDEX 9, France}

\author{Yoichi Ando}
\affiliation{$I\hspace{-.1em}I$. Physikalisches Institut,
Universit\"at zu K\"oln, Z\"ulpicher Str. 77, D-50937 K\"oln,
Germany}

\author{Markus Braden}\email[e-mail: ]{braden@ph2.uni-koeln.de}
\affiliation{$I\hspace{-.1em}I$. Physikalisches Institut,
Universit\"at zu K\"oln, Z\"ulpicher Str. 77, D-50937 K\"oln,
Germany}





\date{\today}

\begin{abstract}

The crystal structure of the candidate topological superconductor \cbs \ was studied by single-crystal neutron diffraction
using samples obtained by inserting the Cu dopant electrochemically. Neither structural refinements nor calculated scattering-density maps find a significant occupation of Cu at the intercalation site between the quintuple layers of Bi$_2$Se$_3$. Following Bragg reflection intensities
as function of temperature, there is no signature of a structural phase transition between 295 and 2\,K. However, the analysis of large sets of Bragg reflections indicates a small structural distortion breaking the rotational axis due to small displacements of the Bi ions.

\end{abstract}


\maketitle


\section{Introduction}

The discovery of superconductivity in Cu-intercalated Bi$_2$Se$_3$ \cite{Hor2010} attracts strong interest, because these
materials are proposed to be candidates for topological superconductivity.
Topological superconductivity is caused by a
non-trivial topology in the superconducting wave-function and is expected to lead to novel phenomena and quasiparticles, of which Majorana
fermions are most prominent.
Fu and Berg analyzed the pairing symmetry in \cbs \ and conclude that the strong spin-orbit coupling in doped Bi$_2$Se$_3$
can result in odd-parity superconductivity with such non-trivial topology \cite{Fu2010}.
The spin-orbit coupling circumvents a strict separation in singlet and triplet pairing, and the spin-orbit-coupled internal degree of freedom takes a triplet.
Besides through Cu intercalation, Bi$_2$Se$_3$ exhibits superconductivity also for Sr \cite{Liu2015,Shruti2015} or Nb \cite{Qiu2015} insertion.
Support for unconventional superconductivity with spin-triplet like pairing is detected in temperature dependent NMR Knight shift experiments, which find only a small change
below $T_c$ in the electronic spin susceptibility for most directions of the magnetic field parallel to the layers \cite{matano2016}.
This experiment also observes a two-fold axis of the Knight shift in the superconducting state, which indicates
that the rotational threefold symmetry is broken below $T_c$ \cite{matano2016}.
The superconducting transition can thus be associated with nematicity.
In the meantime, there is strong support that the superconducting state in all three doped Bi$_2$Se$_3$ systems exhibits a lower symmetry \cite{Yonezawa2018}.
In this nematic superconductivity the threefold rotation axis of the parent material is broken most likely by the anisotropy of the gap amplitude (not only the phase) \cite{Yonezawa2018}.
Critical magnetic fields exhibit a huge in-plane anisotropy following only a two-fold axis \cite{Yonezawa2017}.
Also, the magnetization, resistivity and magnetoresistance exhibit two-fold symmetry leading to the conclusion that the nematicity is
essential for the superconducting phase in doped Bi$_2$Se$_3$ \cite{Asaba2017,Pan2016,Du2017,Shen2017,Smylie2018}.

\begin{figure}
\includegraphics[width=0.9\columnwidth]{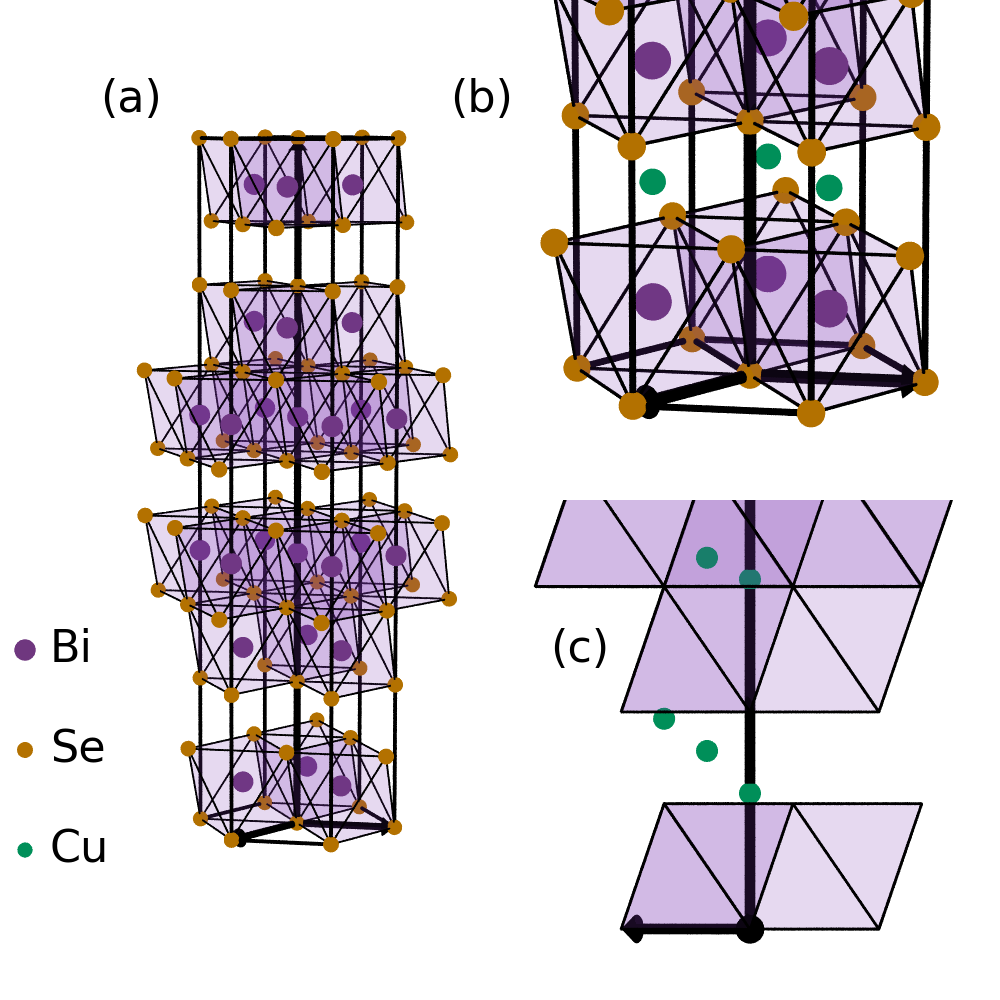}
   \caption{\label{struct} (a) Crystal structure of Bi$_2$Se$_3$ showing the quintuple layers of Bi$_2$Se$_3$ separated by the van-der-Waals gaps. (b) Proposed crystal structure of Cu-doped Bi$_2$Se$_3$ with Cu atoms intercalated between the quintuple layers at $(0, 0, 1/2)$ as proposed in reference  \onlinecite{Hor2010}.
The Cu sites are expected to be not fully occupied but to depend on the Cu content $x$. (c) Possible positions for the Cu atoms $ \mathrm{Cu(i)} $ to $ \mathrm{Cu(v)} $ according to reference  \onlinecite{Wang2011}.}
  \label{disp}
 \end{figure}

An improvement of the sample preparation by electrochemical intercalation and by an annealing step considerably raised the shielding fraction in \cbs \cite{Kriener2011}.
These improved samples exhibit a well-defined anomaly in the specific heat that documents the bulk nature of the superconductivity.
Superconductivity occurs in a broad doping range of 0.1$\le$$x$$\le$0.5 and the transition temperature varies less sharply with
doping than the shielding fraction\cite{Kriener2011,Kriener2011a}.
Furthermore, the charge carrier density does not follow the amount of doping but stays nearly constant at a rather low
value of about 10$^{20}$\,cm$^{-3}$~~\cite{Kriener2011,Kriener2011a}. So far, the three different insertions to induce superconductivity in Bi$_2$Se$_3$
seem to result in essentially the same physical properties, but the Sr and Nb samples are less air-sensitive and superconducting crystals can be grown
from the melt.

The parent compound Bi$_2$Se$_3$ crystallizes in the rhombohedral space group $R\bar{3}m$ and the structure consists of quintuple layers of Bi$_2$Se$_3$ that
are bonded perpendicular to the layers only through the weak van-der-Waals interaction, see Fig. 1 \cite{Nakajima1963}.
In spite of the enormous impact of these candidate topological superconductors, there is little knowledge about the crystal structure
in the doped materials.
Hor et al. deduced from an increase in the $c$ lattice parameter that Cu would occupy an intercalation position in the van-der-Waals gap between two
quintuple layers, at position $ 3b~(0, 0, 1/2) $ see Fig. 1(b) \cite{Hor2010}, but scanning tunnel microscopy, angle resolved photoemission spectroscopy and {\it ab initio} density functional theory proposed other positions - intercalating and interstitial ones - for the dopant \cite{Wang2011}.
Li et al. report a structural analysis of Sr$_x$Bi$_2$Se$_3$ by high-resolution transmission electron microscopy and DFT calculations concluding that the Sr dopant is not occupying the intercalation position \cite{Li2018}.  In addition, a small structural distortion in the Sr$_x$Bi$_2$Se$_3$ lattice of about 0.02\% was deduced from a high resolution X-ray study of d-values at ambient temperature \cite{Kuntsevich2018}.

Here we report on neutron diffraction studies on superconducting \cbs \ crystals at room temperature, as well as at temperatures slightly above and below the superconducting transition.
Cu-doped Bi$_2$Se$_3$ is the prototype compound of this family of unconventional superconductors and such samples are used in various experiments including those documenting the nematic character for the first time. Therefore, we consider it most important to also clarify the crystal structure and the position of the dopant for this material. We cannot detect Cu occupation at any of the proposed intercalation and interstitial sites, and there is no indication for a structural phase transition
between room temperature and 2\,K. However, structural refinements with large data sets improve when the symmetry is lowered to monoclinic.

\section{Experimental}

$ \mathrm{Bi_2Se_3} $ single crystals were grown from a stoichiometric melt. Pieces of the single crystals were electrochemically doped by $ \mathrm{Cu} $ using a saturated solution of $ \mathrm{CuI} $ in $ \mathrm{CH_3CN} $. Typical stoichiometries for superconducting $ \mathrm{Cu}_x\mathrm{Bi_2Se_3} $ lie in the range $ 0.12 \le x \le 0.6 $, and shielding fractions of up to $ 50\,\% $ can be achieved by this technique but the large crystals required for the neutron diffraction experiment yield lower values. Details of the growth process can be found in reference \onlinecite{Kriener2011a}.

The large samples used for neutron diffraction experiments are shown in Fig. 2. The $ \mathrm{Cu} $ doped samples are sensitive to air; therefore, for storage and transport they need to be sealed in glass tubes. They were cleaved and cut to exhibit a plate-like shape {before the intercalation step} and the direction of smallest extension is the $ c $ direction of the hexagonal cell. Sample S1 exhibits a $ \mathrm{Cu} $ amount of $ x = 0.30(1) $ ($T_c$=3.1\,K), for sample S2 it amounts to $ x = 0.33(1) $ ($T_c$=3.6\,K) and for sample S3 to $ x = 0.31(1) $ ($T_c$=3.4\,K). Throughout the paper numbers in parenthesis give the error bars of the last digits.
Superconducting transition temperatures and shielding fractions were measured in a SQUID magnetometer and the results are shown in Fig. 3. The Cu concentration can be determined
by weighting the crystals during the intercalation step but thus only corresponds to the average over the entire crystal.

 \begin{figure}
\includegraphics[width=\columnwidth]{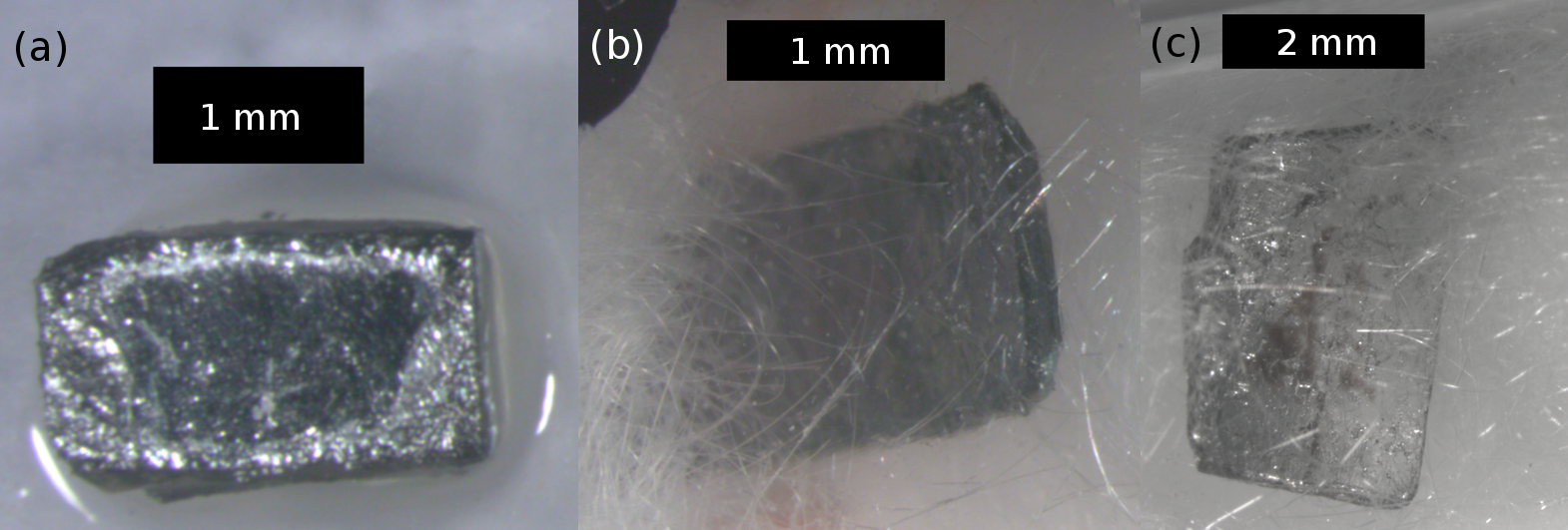}
 \caption{\label{fig:CuxBi2Se3_samples}$ \mathrm{Cu}_x\mathrm{Bi_2Se_3} $ samples for neutron diffraction. (a) S1, (b) S2, (c) S3. Since the samples are air sensitive, the photos of the last two samples were taken with the samples sealed in glass tubes and mechanically protected by quartz wool.}
  \label{disp}
 \end{figure}

\begin{table}
\begin{tabular}{c c c c c c}\hline
\\[-1em]
                        &                      \multicolumn{3}{c}{sample S1} &   \multicolumn{2}{c}{Sample S3}   \\
                T(K)    & 300 & 4.4 & 1.9 & 300 & 2 \\    \hline
  number total          & 530 & 651 & 592 & 640 & 860 \\
 unique $R\overline{3}m$& 157 & 203 & 182 & 137 & 179 \\
 observed $R\overline{3}m$& 124 & 178 & 157 & 557 & 745 \\
 unique $C2/m$          & 515 & 636 & 577 & 129 & 167 \\
 observed $C2/m$        & 358 & 428 & 439 & 459 & 632 \\
\end{tabular}
\caption{\label{tab:numbers} Amount of reflections recorded with the two samples at different temperatures; unique reflections refer to the number of independent
reflections after merging in the corresponding space group, and observed reflections exhibit an intensity to error ratio $\frac{I}{\sigma(I)}>$3.}
\end{table}

Single crystal neutron diffraction was carried out with the three $ \mathrm{Cu}_x\mathrm{Bi_2Se_3} $ crystals  on the single-crystal neutron diffractometer D9 \cite{d9guide,data} at the Institut Laue Langevin using a wavelength of $ 0.84\,\mathrm{\AA}$.
The sample S1 was measured at a temperature of $ 300\,\mathrm{K} $, with a pinhole of $ 5\,\mathrm{mm} $ diameter and at the temperatures of $ 1.9\,\mathrm{K} $ and $ 4.4\,\mathrm{K} $ with a pinhole of $ 6\,\mathrm{mm} $.
During the data collection at $ 300\,\mathrm{K} $, a slight misalignment of the pinhole was detected that however turned out insignificant for the data quality.
At $ 1.9\,\mathrm{K} $ and $ 4.4\,\mathrm{K} $, most of the reflections were collected with exposure times between $ 3\,\mathrm{s} $ and $ 6\,\mathrm{s} $
per point in a combined $\omega$-$n\theta$ scan with $n$ between 0 and 2 \cite{d9guide}; at 300\,K exposure times amounted to between
$ 5\,\mathrm{s} $ and $ 8\,\mathrm{s} $.
With sample S2 only temperature dependent measurements of Bragg intensities were performed. Also with the other two samples Bragg intensities were measured as function of
temperature in order to search for a structural phase transition.
For sample S3,  the optimum pinhole of $ 7\,\mathrm{mm} $ diameter was determined by Renninger scans.
At 300\,K, most of the reflections were collected with exposure times between $ 3\,\mathrm{s} $ and $ 5\,\mathrm{s} $ per data point, and
at $ 2\,\mathrm{K} $ with times between $ 2\,\mathrm{s} $ and $ 7\,\mathrm{s} $.

The intensity data were absorption corrected using the program DATAP.\cite{d9guide} For sample S1, the orientation was exactly known, and a box geometry with a length of $ 2.82\,\mathrm{mm} $, a width of $ 1.55\,\mathrm{mm} $ and a thickness of $ 0.73\,\mathrm{mm} $ was applied. The length and width was determined optically, and the thickness was calculated via the mass of $ 25.23\,\mathrm{mg} $ assuming a density of $ 7.90\,\mathrm{g/cm^3} $ (where the $ \mathrm{Cu} $ doping with $ x = 0.30 $ is taken into account).
Sample S3 exhibits a more irregular plate-like shape and the area was determined optically to amount to $ 8.54\,\mathrm{mm^2} $. Via the mass of $ 46.22\,\mathrm{mg} $ and the density, the thickness of $ 0.69\,\mathrm{mm} $ was determined.

\begin{figure}
\includegraphics[width=0.75\columnwidth]{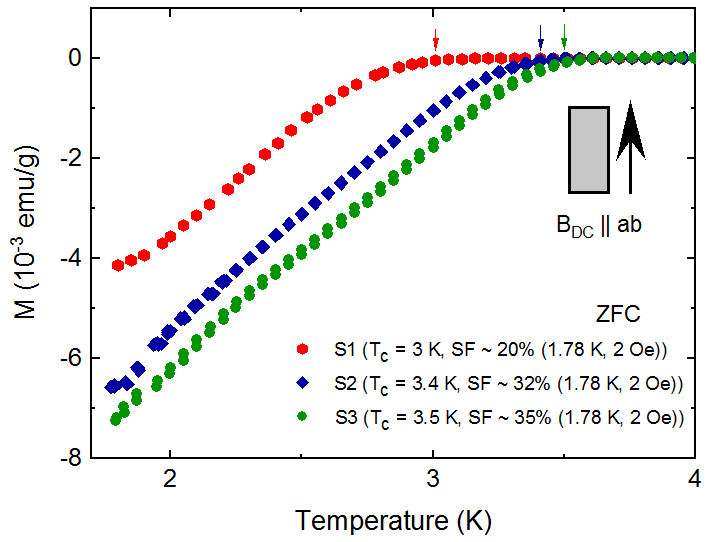}
 \caption{\label{fig:CuxBi2Se3_SC}Zero-field cooled (ZFC) magnetization curves measured with the three single crystals used in the neutron diffraction experiments. The magnetic field of 0.2\,mT was applied along the planes. Shielding fractions (SF) are given in the legend for T=1.78\,K. }
  \label{squid}
 \end{figure}

\section{Results and Discussion}

\subsubsection{Vacancies and possible Cu positions}

First refinements of the crystal structure were performed in space group $ R\overline{3}m $ that is reported for the parent compound and that allows for
various possible Cu positions. Rhombohedral crystals in general can exhibit two twins corresponding to obverse and reverse setting. The refinements of these twinning fractions for sample S1 yield $ -0.99(1.23)\,\% $ at $ 1.9\,\mathrm{K} $, $ -0.02(1.65)\,\% $ at $ 4.4\,\mathrm{K} $ and $ 0.46(23)\,\% $ at $ 300\,\mathrm{K} $ for the reverse setting. The corresponding refinements for sample S3 yield fractions $ 0.39(1.30)\,\% $ at $ 2\,\mathrm{K} $ and $ -0.82(1.99)\,\% $ at $ 300\,\mathrm{K} $ for the reverse setting.  Thus, the samples do not exhibit obverse/reverse twinning.

$ \mathrm{Se} $ vacancies are supposed to play an important role for transport properties in  pure $ \mathrm{Bi_2Se_3} $. Therefore,
the occupation of the two Se sites were refined.
For sample S1 at a temperature of $ 1.9\,\mathrm{K} $, $ 4.4\,\mathrm{K} $ and $ 300\,\mathrm{K} $), the occupation of the $ \mathrm{Se2} $ site is $ 1.011(12) $, $ 0.995(16) $ and $ 1.002(16) $, respectively, and the occupation of the $ \mathrm{Se3} $ site is $ 0.999(11) $, $ 0.998(14) $ and $ 0.985(16) $, respectively.
For sample S3 at 2\,K and 300\,K, the occupation of the $ \mathrm{Se2} $ site is $ 1.008(10) $ and $ 1.003(15) $, respectively, and the occupation of the $ \mathrm{Se3} $ site is $ 0.989(10) $ and $ 1.001(14) $, respectively. These values are consistent with full occupation. Thus, no $ \mathrm{Se} $ vacancies could be detected by neutron diffraction.

These refinements not only exclude $ \mathrm{Se} $ vacancies but also a significant amount  of $ \mathrm{Bi} $ vacancies.
We estimate a possible substitution of $ \mathrm{Bi} $ through $ \mathrm{Cu} $ by their different neutron scattering lengths: $b_{\mathrm{Cu}} = 7.718\,\mathrm{fm}$ and $b_{\mathrm{Bi}}=8.532\,\mathrm{fm}$ \cite{tablesC}. Even though the actual compositions of the crystals are essentially Cu$_x$Bi$_2$Se$_3$, if we hypothesize
 a composition Cu$_x$Bi$_{2-x}$Se$_3$ and apply identical atomic displacement parameters to the Cu and Bi ions, we obtain for sample S1 a Cu substitution of 4(9), -3(12) and -7(13)\,\%  at 1.9, 4.4 and 300\,K, respectively, and for sample S3 a Cu substitution of  -3(8) and 2(11)\,\% at 2 and 300\,K, respectively.
Note that the precision is rather poor due to the small difference between the scattering lengths of Bi and Cu.

\begin{table}
\begin{tabular}{c c c c c c c}\\\hline
 & & $ \mathrm{Cu(i)} $ & $ \mathrm{Cu(ii)} $ & $ \mathrm{Cu(iii)} $ & $ \mathrm{Cu(iv)} $ & $ \mathrm{Cu(v)} $ \\\hline
 \multirow{3}{*}{S1} & $ 1.9\,\mathrm{K} $ & $ 0.4(4) $ & $ 0.1(4) $ & $ 0.6(4) $ & $ 0.4(4) $ & $ -0.37(33) $ \\
                                             & $ 4.4\,\mathrm{K} $ & $ 0.3(5) $ & $ 0.5(5) $ & $ 0.2(5) $ & $ -0.2(5) $ & $ -0.1(4) $ \\
                                             & $ 300\,\mathrm{K} $ & $ 0.4(5) $ & $ 0.3(5) $ & $ 0.4(5) $ & $ 0.2(5) $ & $ 0.0(4) $ \\\hline
 \multirow{3}{*}{S3} & $ 2\,\mathrm{K} $   & $ 0.7(4) $ & $ 0.71(33) $ & $ 0.58(34) $ & $ 0.44(34) $ & $ -0.43(29) $ \\
                                             & $ 300\,\mathrm{K} $ & $ 0.6(5) $ & $ 0.5(5) $ & $ 0.8(5) $ & $ 0.6(5) $ & $ -0.2(4) $ \\\hline
\end{tabular}
\caption{\label{tab:table_CuxBi2Se3_Cu}Occupancies (given in percent of full occupation) of the possible $ \mathrm{Cu} $ positions $ \mathrm{(i)} $ to $ \mathrm{(v)} $ that are all located at a Wyckoff position $6c$ from refinements in space group $ R\overline{3}m $. There is no significant deviation from zero. Each value was determined by a separate refinement without considering the other $ \mathrm{Cu} $ positions.}
\end{table}

$ \mathrm{Cu} $ intercalation in the van-der-Waals gap is expected to enhance the thickness of the van-der-Waals gap and thus to cause
an increase of the $c$ lattice parameter. For pristine $ \mathrm{Bi_2Se_3} $, the lattice constants $ c = 28.636(20)\,\mathrm{\AA} $ \cite{Nakajima1963}, $28.615(2)\,\mathrm{\AA} $ \cite{Vicente1999} and $ 28.666(1)\,\mathrm{\AA} $ \cite{Hor2010} were reported and for $ \mathrm{Cu_{0.12}Bi_2Se_3} $, the parameter increases
to 28.736(1)\,\AA\, \cite{Hor2010}.
LeBail refinements of $ (0\,0\,l)$ scans performed on a D5000 X-ray diffractometer with our single crystals
yield a lattice constant $ c = 28.5990(8)\,\mathrm{\AA} $ for  undoped $ \mathrm{Bi_2Se_3} $ and 28.683(2)\,\AA \ for $ \mathrm{Cu_{0.3}Bi_2Se_3} $
confirming the increased lattice constant $c$ due to the insertion of Cu, $\Delta c$=0.084\AA\, $\sim2.9\cdot10^{-3}c$.

The occupation of the most obvious intercalation position at $ 3b~(0, 0, 1/2) $ was examined with the five data sets obtained on two distinct
samples by comparing refinements with full occupation, with partial occupation and without any Cu at this position.
For the $ \mathrm{Cu} $ atom, the displacement parameter was fixed to isotropic $ U = 0.005\,\mathrm{\AA^2} $ for the high temperature $ 300\,\mathrm{K} $ respectively $ U = 0.0025\,\mathrm{\AA^2} $ for the low temperatures $ 1.9\,\mathrm{K} $, $ 2\,\mathrm{K} $ and $ 4.4\,\mathrm{K} $.
The results are shown
in Table \ref{tab:occupations} in the Appendix. The five data sets consistently exclude a significant presence of Cu at this position.  By
combining scanning tunnel microscopy, angle resolved photoemission spectroscopy and {\it ab initio} density functional theory, Wang et al. conclude that the most probable sites for the $ \mathrm{Cu} $ atoms are two interstitial sites whithin the quintuple layers and three intercalated sites between the quintuple layers \cite{Wang2011}. From the caption of Fig. 4 in reference \cite{Wang2011}, it is possible to reconstruct these coordinates. Using the structural parameters from reference \onlinecite{Nakajima1963}, one gets the following positions: The interstitial atom $ \mathrm{Cu(i)} $ is located at the position $ 6c~(1/3, 2/3, 0.36127(34)) $, and the other interstitial position $ \mathrm{Cu(ii)} $ is $ 6c~(0, 0, 0.34032(33)) $. The coordinates of the three intercalated atoms are: $ 6c~(0, 0, 0.1321(15)) $ for $ \mathrm{Cu(iii)} $, $ 6c~(2/3, 1/3, 0.2047(15)) $ for $ \mathrm{Cu(iv)} $ and $ 6c~(1/3, 2/3, 0.1733(15)) $ for $ \mathrm{Cu(v)} $.
Other possibilities can be found in references \onlinecite{Sobczak2018a} and \onlinecite{Li2018}. All positions proposed in reference \onlinecite{Wang2011} are tested via structural refinements.
A significant occupation was detected in none of these positions, see table I.

Symmetrized Fourier maps were calculated for the refined models in space group $ R\overline{3}m $ for sample S1 at $ 1.9\,\mathrm{K} $  and for sample S3 at $ 2\,\mathrm{K} $, see Fig. 7 in the Appendix, using the software JANA2006 \cite{petricek}. The maps show the nuclear scattering density within the unit cell. If an atom is missing in the refinement, there should be an additional peak at this position. All atoms of $ \mathrm{Bi_2Se_3} $ are clearly visible but no other features appear in the Fourier maps. Shallow peaks like the one at $ (1/3, 2/3, 0.025) $ appearing vertically displaced from occupied positions might point to some occupational mixing or insufficient treatment of the
atomic displacements. The Fourier maps thus  do not indicate the location of the extra Cu ions, and in particular they confirm the absence of a significant amount of Cu at $ (0, 0, 1/2) $.

With the precise $c$ lattice parameters determined by our X-ray diffraction experiment and with the structural parameters of reference \onlinecite{Vicente1999} we can compare the
width of the van-der-Waals gap, which amounts to 2.532(4)\,\AA \, for the parent compound. For the doped compound, we use the average of the $z$ parameter of $Se3$ determined at
300\,K and obtain 2.565(2)\,\AA. The enhancement of the $c$ lattice parameter is thus realized by an increase of the van-der-Waals gap giving support to the idea that
Cu ions are inserted in this gap.

\begin{figure}
\includegraphics[width=0.95\columnwidth]{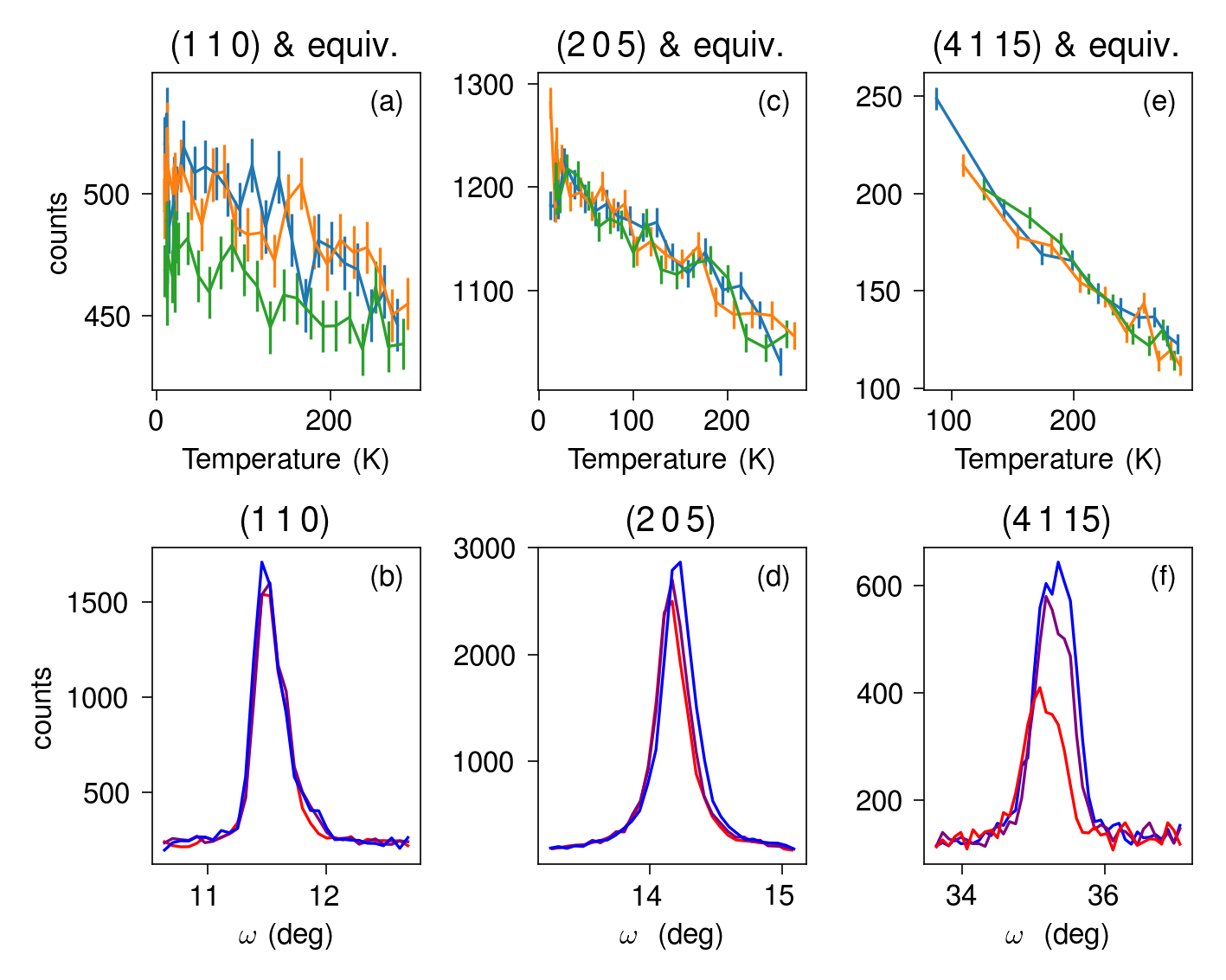}
  \caption{\label{reflections} The upper panels show the temperature dependence of the integrated intensities
  for three set of equivalent reflections in $R\overline{3}m$: panel (a) (1 1 0), (-1 2 0) and (-2 1 0), panel (b) (2 0 5), (-2 2 5) and (0 -2 5),
  panel (c) (4 1 15), (1 -5 15) and (-5 4 15). The lower panels show rocking-scan profiles for three different
  temperatures (the highest temperature in red, the middle in violet and the lowest in blue). In panel (a) the (1 1 0) is shown
  at 277, 156 and 9\,K; in panel (b) (2 0 5) at 255, 158 and 12\,K; in panel (c) (4 1 15) at 284, 143 and 88\,K.
 }
  \label{refl-t}
\end{figure}

It is thus not possible to detect the position of the inserted Cu with our comprehensive diffraction experiments. At the initially proposed intercalation position and at Cu (v) no Cu can be detected at all, and also the other positions Cu (i) to (iv) only yield a total contribution of 0.036(10) \cite{note-calc} for $x$ in the formula \cbs\, while $x\sim0.3$ is expected. We must conclude that the Cu atoms do not occupy a well defined position in \cbs , which otherwise would have been detected in the Fourier maps. Instead there must be some positional smearing, which results in very large atomic displacement parameters impairing the detection of the dopant by diffraction. Part of this positional smearing can stem
from the separation of superconducting and non-superconducting regions that should differ in their Cu content in the distribution of the dopants in the lattice.
The ill-defined dopant position can furthermore be the consequence of clustering.

\subsubsection{Structural phase transition and monoclinic distortion}

If a structural phase transition occurs between room and low temperature, one expects some anomalies in the temperature dependence of
fundamental Bragg peaks as well. The intensities of three different reflections and their equivalents in space group $ R\overline{3}m $ were followed upon cooling, see Fig. 4.
The integrated intensities of most equivalent reflections agree even at low temperature, and there is no anomaly visible in the temperature dependencies of the
integrated intensities, which reflect the smaller atomic displacement parameters at low temperature. That there is an enhanced temperature dependence
in the reflections (4 1 15) and equivalent ones most likely arises from the different temperature dependencies of the quintuple layers and the van-der-Waals gaps, but again there
is no evidence for a stuctural phase transition.
Also the width of the reflections do not yield any indication for structural phase transition, as the peak profiles that are shown for three reflections in the lower panels
of Fig. 4 do not change.
These temperature-dependent data were taken with $\omega$ scans and a two-dimensional detector resulting in three-dimensional data sets for each Bragg peak at each temperature.
We have numerically analyzed the resemblance of these three-dimensional data taken at different temperatures for an individual Bragg peak, see Fig. \ref{fig:correlation} in the Appendix. This procedure
would also detect some anomaly in the mosaic spread due to twining, but there is no indication for any anomaly in this resemblance analysis.
The absence of a structural phase transition agrees with specific-heat measurements indicating a smooth temperature dependence for
Sr-doped \cite{Pan2016,Willa2018} and for Nb-doped \cite{Asaba2017} Bi$_2$Se$_3$.

\begin{figure}
\includegraphics[width=0.78\columnwidth]{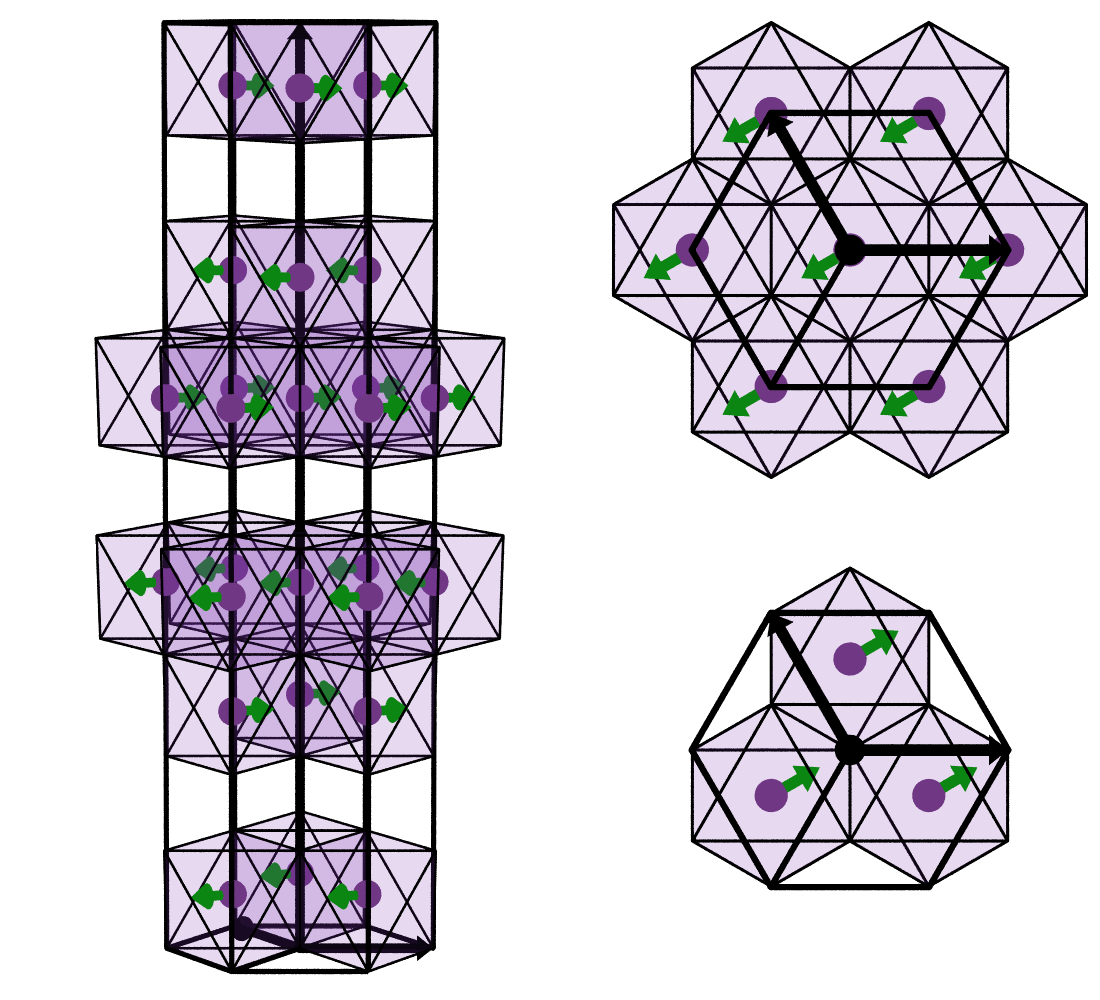}
  \caption{\label{structure} Left:
 Distorted crystal structure of \cbs \ as described in space group $C2/m$. The displacement of the Bi positions of the refinements in $ C2/m $ compared to the positions in $ R\overline{3}m $ are displayed by arrows. The arrows are $50$ times as large as the actual distortions. Right: Two neighboring quintuple layers (view along the negative $ c $ axis).
  The displacements are within the monoclinic $a,c$ plane, the mirror plane of $ C2/m $.}
  \label{disp}
 \end{figure}

\begin{table}
\begin{tabular}{c c c c c c c c}\hline
\\[-1em]
                                      &                                       &       &\multicolumn{2}{c}{$ R\overline{3}m $} &  &  \multicolumn{2}{c}{$ C2/m $}   \\
                                      &                                       &       &    Bi1 6c         & Se3 6c           &  & Bi1 4i       & Se3 4i      \\\cline{2-5}\cline{7-8}
 \multirow{9}{*}{\rotatebox{90}{sample: S1}}  & \multirow{3}{*}{$ 1.9\,\mathrm{K} $}  & $ x $ & $ 0 $             & $ 0 $     &  & $ -0.0095(5) $  & $ -0.0010(9) $ \\
                                      &                                       & $ y $ & $ 0 $             & $ 0 $     &  & $ -0.0048(3) $  & $ -0.0005(4) $ \\
                                      &                                       & $ z $ & $ 0.40068(4) $             & $ 0.21098(5) $     &  & $ 0.40069(3) $  & $ 0.21092(3) $ \\\cline{2-5}\cline{7-8}
                                      & \multirow{3}{*}{$ 4.4\,\mathrm{K} $}  & $ x $ & $ 0 $             & $ 0 $     &  & $ 0.0088(7) $  & $ 0.0035(11) $ \\
                                      &                                       & $ y $ & $ 0 $             & $ 0 $     &  & $ 0.0044(4) $  & $ 0.0017(5) $ \\
                                      &                                       & $ z $ & $ 0.40064(6) $             & $ 0.21076(6) $     &  & $ 0.40064(4) $  & $ 0.21072(4) $ \\\cline{2-5}\cline{7-8}
                                      & \multirow{3}{*}{$ 300\,\mathrm{K} $}  & $ x $ & $ 0 $             & $ 0 $     &  & $ -0.0103(11) $  & $ -0.0021(15) $ \\
                                      &                                       & $ y $ & $ 0 $             & $ 0 $     &  & $ -0.0052(5) $  & $ -0.0010(7) $ \\
                                      &                                       & $ z $ & $ 0.40061(7) $             & $ 0.21126(7) $     &  & $ 0.40060(5) $  & $ 0.21121(5) $ \\\hline
                                      & \\\hline
\\[-1em]

                                      &                                       &       &\multicolumn{2}{c}{$ R\overline{3}m $} &  &  \multicolumn{2}{c}{$ C2/m $}   \\
                                      &                                       &       &     Bi1               & Se3           &  & Bi1        & Se3       \\\cline{2-5}\cline{7-8}
 \multirow{6}{*}{\rotatebox{90}{sample: S3}}  & \multirow{3}{*}{$ 2\,\mathrm{K} $}    & $ x $ & $ 0 $             & $ 0 $     &  & $ -0.0078(3) $  & $ -0.0001(5) $ \\
                                      &                                       & $ y $ & $ 0 $             & $ 0 $     &  & $ -0.00392(17) $  & $ -0.0001(2) $ \\
                                      &                                       & $ z $ & $ 0.40067(4) $             & $ 0.21112(4) $     &  & $ 0.40070(2) $  & $ 0.21113(2) $ \\\cline{2-5}\cline{7-8}
                                      & \multirow{3}{*}{$ 300\,\mathrm{K} $}  & $ x $ & $ 0 $             & $ 0 $     &  & $ -0.0120(7) $  & $ -0.0009(9) $ \\
                                      &                                       & $ y $ & $ 0 $             & $ 0 $     &  & $ -0.0060(3) $  & $ -0.0004(4) $ \\
                                      &                                       & $ z $ & $ 0.40070(7) $             & $ 0.21150(7) $     &  & $ 0.40078(4) $  & $ 0.21148(4) $ \\\hline
\end{tabular}
\caption{\label{tab:positional_parameters}Positional parameters of the refinements in space groups $ R\overline{3}m $ and $ C2/m $. Both structures are described here with respect to the hexagonal axes.
In both structures Se2 is located at the origin (Wyckoff position $ 3a~0,0,0 $ and $ 2a~0,0,0 $, respectively). The lattice parameters amount to $ a = b = 4.143(5)\,\mathrm{\AA} $ \cite{Nakajima1963} and $ c = 28.683(2)\,\mathrm{\AA} $ at 300\,K. The $ c $ lattice constant was determined by a single-crystal X-ray $ (0\,0\,l) $ scan.}
\end{table}

Breaking of translation symmetry can give rise to superstructure reflections that are forbidden in space group $ R\overline{3}m $ and break the general selection rule $ -h + k + l = 3n $ \cite{tablesA}.
$ 152 $ such forbidden reflections were collected at $ 300\,\mathrm{K} $ and $ 167 $ at $ 1.9\,\mathrm{K} $ with sample S1. If these reflections are absent, it is expected that
$\sim$68\,\% \ of the experimental values lie within one standard deviation away from $0$. More specifically: The quotient $\frac{I}{\sigma(I)}$ of the intensity divided by its error should exhibit a Gauss distribution with mean value $ 0 $ and standard deviation $ 1 $. For a few forbidden reflections, a neighboring strong Bragg reflection induces
significant intensity, but inspection of the three-dimensional data shows an off-centering and thus a contamination of neighboring Bragg peaks due to the
long $c$ axis.
After excluding such contaminated reflections
the entity of the normalized forbidden intensities exhibits the expected Gauss distribution centered at zero with width one; therefore we can exclude a significant
breaking of the translational symmetry in \cbs . The same conclusion was also drawn for Sr-doped Bi$_2$Se$_3$ \cite{Smylie2018}.

However, the symmetry can be lower than $R\overline{3}m$ and keep the full translation symmetry, which
means that the structural distortion is described by an irreducible representation at the $\Gamma$ point of the Brillouin zone\cite{stokes}.
There are only two isotropy subgroups of $R\overline{3}m$ that are centrosymmetric and break the threefold rotation axis, $C2/m$ and $P\overline{1}$,\cite{stokes} and these two
only differ in the image of the multidimensional order parameter. Therefore, we may restrict the analysis to the simpler one,  $C2/m$, but need to
take twinning into account. The monoclinic lattice results from the rhombohedral one by $a'=-2/3a-1/3b+2/3c$, $b'=-b$ and $c'=2/3a+1/3b+1/3c$ without a
shift of the origin. Note that the primitive lattices are identical for $R\overline{3}m$ and $C2/m$ and there no additional selection rules. 

In space group $C2/m$, as in $R\overline{3}m$, there are also three inequivalent sites, Bi1, Se2 and Se3,
and  compared to the high-symmetry refinement, there are only two additional position parameters, an in-plane displacement for Bi1 and for Se3.
For the description of the atomic displacement parameters we keep the constraints from space group $R\overline{3}m$: $U_{11}$=$U_{22}$, $U_{13}$=$U_{23}$=0 and
$U_{12}$=$2U_{11}$. Therefore the refinement in space group $C2/m$ adds only four additional parameters (two positional parameters and two twin fractions).
We can compare the reliability factors of the $C2/m$ refinements with those obtained in $R\overline{3}m$ by describing the high symmetry phase in the low-symmetry
setting. As can be seen in Table \ref{tab:CuxBi2Se3_Rvalues_3axis} in the Appendix, for the three refinements there is a significant improvement in all reliability factors that is also reflected in a reduction of the goodness of fit values.
Note that the $R\overline{3}m$ and $C2/m$ refinements cannot be directly compared, because in the former case the data set is much smaller by merging many equivalent
reflections.

Further support for the symmetry reduction is given by the good agreement of the five
different refinements. We always find a significant displacement of the Bi positions while the Se3 ion stays at the high symmetry position as shown in Table \ref{tab:positional_parameters}. The Bi displacement is illustrated in Fig. 5.
The twin fractions indicate an almost equal occupation of the three twin orientations in the two large crystals studied here.
Note, however, that the mono\-clinic distortion of the translation lattice is too small to be detected with the resolution of the
single-crystal neutron diffraction experiment.

In spite of its weakness the structural distortion can be essential for the
observation of a large anisotropy in the nematic superconducting state by poling the superconducting domains similar to the case of Cu$_{1.5}$(PbSe)$_5$(Bi$_2$Se$_3$)$_6$ where the
nematic axis follows an intrinsic structural symmetry breaking \cite{Andersen2018}.
The monoclinic distortion is the natural candidate to generate the uniaxial strain discussed by Venderbos et al. \cite{Venderbos2016} to lift
the degeneracy of the two superconducting components. The nematic superconducting transition exhibits an improper ferroelastic character and
can thus be efficiently poled by small strain.

Evidence for a structural distortion was obtained also for Sr-doped Bi$_2$Se$_3$ by high resolution X-ray experiments at room temperature \cite{Kuntsevich2018} and by
a tiny anomaly in the thermal expansion occurring slightly above the superconducting transition in a Nb doped sample \cite{Cho2019}. In both cases the structural distortion
are well below the resolution in our experiments. But qualitatively the room-temperature distortion reported for the Sr system \cite{Kuntsevich2018} corresponds
to our finding of a monoclinic space group for Cu doping.

Further support for the reliability of these structural refinements can be obtained from the anisotropic displacement parameters given in Table V in the Appendix.
The small displacement of the Bi atom is visible in a minor reduction of the Bi $U_{11}$ parameter in space group $C2/m$. In general there is a strong
anisotropy of the displacement parameters with $U_{33}>U_{11}$, which can be explained by the layered character of the \cbs \, structure that exhibits strong
bonding only within the quintuple layers. Therefore, $c$ polarized phonons must be much softer and imply large $U_{33}$ values.

In the high-symmetry space group $ R\overline{3}m $, there are two different $ \mathrm{Bi-Se} $ distances: One from $ \mathrm{Bi} $ to the $ \mathrm{Se} $ atom at the border of the van-der-Waals gap and another one to the $ \mathrm{Se} $ in the middle of the quintuple layer. These two distances amount to $ 2.8642(25)\,\mathrm{\AA} $ and $ 3.0745(23)\,\mathrm{\AA} $, respectively. We use the $c$ parameters obtained on our crystals, the in-plane lattice parameter from reference \cite{Nakajima1963} and the structural parameters are averaged from refinements of the data collected at $ 1.9\,\mathrm{K} $ and $ 2\,\mathrm{K} $. In the low-symmetry space group $ C2/m $, each of these two distances splits into two different ones, see Fig. 5. The distance from the $ \mathrm{Bi} $ atom to one of the the border $ \mathrm{Se} $ atoms amounts to $ 2.8381(28)\,\mathrm{\AA} $ and the distances to the other two $ \mathrm{Se} $ atoms at the border are $ 2.8769(26)\,\mathrm{\AA} $. The other distance splits into $ 3.0980(24)\,\mathrm{\AA} $ (occurring once) and two $ 3.0635(23)\,\mathrm{\AA} $ distances.
Already in the high-symmetry phase the Bi ion is not situated at the center of the surrounding octahedron but moves towards the border of the quintuple layer.
In Bi$_2$Se$_3$, Bi shows a formal 3+ valence state, which typically exhibits lone-pair behavior \cite{orgel1959,galy1975,laarif1986}. For Bi$^{3+}$, there is a pair of electrons outside the filled Xe shell, whose probability density can deviate from spherical symmetry. The strong off centering in the high-symmetry structure can already be related
with such an effect. The structural symmetry-breaking displacement in the low-symmetry phase $ C2/m $ allows for horizontal offcentering of the lone-pair electron
distribution.

If one admits a loss of inversion symmetry, the low-symmetry phase breaking the rotational axis without a loss of translation symmetry corresponds to space group $Cm$ \cite{stokes}. Note, however, that there is no evidence of broken inversion symmetry so far.
Describing the structure in $Cm$ requires 13 additional structural parameters and at least two twin parameters. We have refined this structural model
with all 5 data sets, but we do not obtain evidence for such a distortion. The large number of additional parameters combined with the twinning impairs a reliable refinement.

\section{Conclusions}

Our comprehensive neutron diffraction studies on superconducting single crystals of Cu$_x$Bi$_2$Se$_3$ with $x\sim 0.3$ show that the dopant is not simply occupying the central intercalation position, but also at other proposed sites we do not detect significant amounts of Cu. The position of the Cu ion remains a mystery, but the enhanced
thickness of the van-der-Waals gap supports the idea of intercalation between the layers. For any Cu
incorporated in the Bi$_2$Se$_3$ structure, we assume considerable disorder arising from clustering to hide it in diffraction studies. We also do not find any
evidence for a structural phase transition occurring between room temperature and the superconducting phase similar to caloric studies. However, the five collected Bragg reflection data sets
consistently indicate a weak structural distortion lowering the symmetry to the monoclinic space group $ C2/m $. This monoclinic distortion
can be related to the lone electron pair of the Bi and be essential for the analysis of the nematic superconductivity in doped Bi$_2$Se$_3$ in spite of its weakness.

\begin{acknowledgments}
This work was funded by the Deutsche Forschungsgemeinschaft (DFG,
German Research Foundation) - Project number 277146847 - CRC 1238, projects A02, A04 and  B04.
\end{acknowledgments}

\section{Appendix}

\subsection{Additional information about the refinements}

In table III, we compare the reliability factors of refinements in space group $ R\overline{3}m $ with empty, full and partially occupied intercalation position at (0 0 1/2).
We can exclude a significant Cu occupation of this site.

\begin{table}[h]
\begin{tabular}{c c r r r r r r}\hline
                                      & \multicolumn{2}{l}{Sample: S1} \\
                                      &                                       & \multicolumn{2}{c}{without $ \mathrm{Cu} $} &  \multicolumn{2}{c}{with $ \mathrm{Cu} $} &  \multicolumn{2}{c}{free $\mathrm{occ(Cu)} $} \\\hline
                                      & \multirow{3}{*}{$ 1.9\,\mathrm{K} $}  & $ R_\mathrm{obs}(\mathrm{w}R_\mathrm{obs}) : $  & $ 3.07(3.55)  $ &      & $ 12.99(14.48)  $  &  $ 3.07(3.54)  $\\
                                      &                                       & $ R_\mathrm{all}(\mathrm{w}R_\mathrm{all}) : $  & $ 3.71(3.71)  $ & $  $ & $ 14.09(14.61)  $  &  $ 3.69(3.70)  $\\
                                      &                                       &                       &              &                         &         $ \mathrm{occ(Cu)} : $  & $ -0.006(5)     $\\\cline{2-8}
                                      & \multirow{3}{*}{$ 4.4\,\mathrm{K} $}  & $ R_\mathrm{obs}(\mathrm{w}R_\mathrm{obs}) : $  & $ 5.05(5.25)  $ & $  $ & $ 14.51(16.25)  $  &  $ 4.98(5.24)  $\\
                                      &                                       & $ R_\mathrm{all}(\mathrm{w}R_\mathrm{all}) : $  & $ 5.71(5.36)  $ & $  $ & $ 15.36(16.34)  $  &  $ 5.74(5.38)  $\\
                                      &                                       &                       &               &                         &        $ \mathrm{occ(Cu)} : $  & $ 0.000(7)     $\\\cline{2-8}
                                      & \multirow{3}{*}{$ 300\,\mathrm{K} $}  & $ R_\mathrm{obs}(\mathrm{w}R_\mathrm{obs}) : $  & $ 3.15(4.83)  $ & $  $  & $ 15.32(17.29) $  &  $ 3.15(4.83)  $\\
                                      &                                       & $ R_\mathrm{all}(\mathrm{w}R_\mathrm{all}) : $  & $ 4.69(5.04)  $ & $   $ & $ 16.92(17.40)  $ &  $ 4.69(5.04)  $\\
                                      &                                       &                       &               &                         &        $ \mathrm{occ(Cu)} : $  & $ 0.000(7)     $\\\hline
                                      &                                       &                       &               &                         &             &                            &              \\\hline
                                      & \multicolumn{2}{l}{Sample: S3} \\
                                      &                                       & \multicolumn{2}{c}{without $ \mathrm{Cu} $} &  \multicolumn{2}{c}{with $ \mathrm{Cu} $} &  \multicolumn{2}{c}{free $\mathrm{occ(Cu)} $} \\\hline
                                      & \multirow{3}{*}{$ 2\,\mathrm{K} $}    & $ R_\mathrm{obs}(\mathrm{w}R_\mathrm{obs}) : $  & $ 3.12(3.97)  $ & $  $  & $ 14.49(18.97)  $ &  $ 3.10(3.96)  $\\
                                      &                                       & $ R_\mathrm{all}(\mathrm{w}R_\mathrm{all}) : $  & $ 3.24(3.99)  $ & $   $ & $ 14.78(19.02)  $ &  $ 3.21(3.97)  $\\
                                      &                                       &                       &               &                         &        $ \mathrm{occ(Cu)} : $  & $ -0.006(5)     $\\\cline{2-8}
                                      & \multirow{3}{*}{$ 300\,\mathrm{K} $}  & $ R_\mathrm{obs}(\mathrm{w}R_\mathrm{obs}) : $  & $ 4.56(5.08)  $ & $  $  & $ 14.87(17.31)  $ &  $ 4.54(5.07)  $\\
                                      &                                       & $ R_\mathrm{all}(\mathrm{w}R_\mathrm{all}) : $  & $ 4.83(5.11)  $ & $   $ & $ 15.54(17.35)  $ &  $ 4.79(5.10)  $\\
                                      &                                       &                       &               &                         &        $ \mathrm{occ(Cu)} : $  & $ -0.005(7)     $\\\hline
\end{tabular}
\caption{\label{tab:occupations} Occupation of the intercalation position (0 0 1/2) by Cu in the refinements in space group $R\overline{3}m$; all reliability values are given in  \% .}
\end{table}

Table IV compares the reliability factors of the refinements in space groups  $R\overline{3}m$ and $C2/m$. The column below $R\overline{3}m$ gives the standard
values for the high-symmetry refinement, which, however, cannot be directly compared to those in space group $C2/m$ because data are differently merged. Therefore, we
have also described the high-symmetry phase in space group $C2/m$ by restraining the two extra parameters and by setting fixed domain ratios. This can be
directly compared to the refinements in $C2/m$ as indicated by the arrows. With the five data sets, we obtain a weak but significant reduction of the reliability
parameters in agreement with improved goodness of fit values.

\begin{table}
\begin{tabular}{c c r r r r}\hline
                                      & \multicolumn{2}{l}{Sample: S1} \\
                                      &                                       & \multicolumn{2}{c}{$ R\overline{3}m $} & \multicolumn{2}{c}{$ C2/m $}   \\\hline
                                      & \multirow{4}{*}{$ 1.9\,\mathrm{K} $}  & $ R_\mathrm{obs}(\mathrm{w}R_\mathrm{obs}) = $  & $ 3.07(3.55)  $ & $  $  & $ 3.35(3.69)  \rightarrow  3.07(3.46)  $ \\
                                      &                                       & $ R_\mathrm{all}(\mathrm{w}R_\mathrm{all}) = $     & $ 3.71(3.71)  $ & $   $  & $ 4.82(4.04)  \rightarrow  4.68(3.62)  $ \\
                                      &                                       & $ \mathrm{GoF} = $                  & $ 2.07 $     &    & $ 1.67 \rightarrow  1.51 $ \\\cline{2-6}
                                      & \multirow{4}{*}{$ 4.4\,\mathrm{K} $}  & $ R_\mathrm{obs}(\mathrm{w}R_\mathrm{obs}) = $     & $ 5.05(5.25)  $ & $   $  & $ 4.90(5.33)  \rightarrow  4.78(5.22)  $ \\
                                      &                                       & $ R_\mathrm{all}(\mathrm{w}R_\mathrm{all}) = $     & $ 5.71(5.36)  $ & $   $  & $ 6.57(5.69)  \rightarrow  6.38(5.36)  $ \\
                                      &                                       & $ \mathrm{GoF} = $                  & $ 2.81 $     &    & $ 2.21 \rightarrow  2.10 $ \\\cline{2-6}
                                      & \multirow{4}{*}{$ 300\,\mathrm{K} $}  & $ R_\mathrm{obs}(\mathrm{w}R_\mathrm{obs}) = $     & $ 3.15(4.83)  $ & $   $  & $ 4.07(5.29)  \rightarrow  3.97(5.22)  $ \\
                                      &                                       & $ R_\mathrm{all}(\mathrm{w}R_\mathrm{all}) = $     & $ 4.69(5.04)  $ & $   $  & $ 6.32(5.67)  \rightarrow  6.29(5.47)  $ \\
                                      &                                       & $ \mathrm{GoF} = $                  & $ 2.74 $     &    & $ 2.31 \rightarrow  2.25 $ \\\hline
                                      & \\\hline
                                      & \multicolumn{2}{l}{Sample: S3} \\
                                      &                                       & \multicolumn{2}{c}{$ R\overline{3}m $}  & \multicolumn{2}{c}{$ C2/m $}   \\\hline
                                      & \multirow{4}{*}{$ 2\,\mathrm{K} $}    & $ R_\mathrm{obs}(\mathrm{w}R_\mathrm{obs}) = $   & $ 3.12(3.97)  $ & $   $   & $ 3.41(3.99)  \rightarrow 2.96(3.52)  $ \\
                                      &                                       & $ R_\mathrm{all}(\mathrm{w}R_\mathrm{all}) = $                 & $ 3.24(3.99)  $ & $   $  & $ 4.14(4.14)  \rightarrow 3.79(3.60)  $ \\
                                      &                                       & $ \mathrm{GoF} = $                  & $ 2.77 $     &   & $ 2.16 \rightarrow  1.90 $ \\\cline{2-6}
                                      & \multirow{4}{*}{$ 300\,\mathrm{K} $}  & $ R_\mathrm{obs}(\mathrm{w}R_\mathrm{obs}) = $   & $ 4.56(5.08)  $ & $   $  & $ 4.78(5.43)  \rightarrow 4.24(4.94)  $ \\
                                      &                                       & $ R_\mathrm{all}(\mathrm{w}R_\mathrm{all}) = $   & $ 4.83(5.11)  $ & $   $  & $ 5.83(5.60)  \rightarrow 5.68(5.03)  $ \\
                                      &                                       & $ \mathrm{GoF} = $                  & $ 3.26 $     &   & $ 2.55 \rightarrow  2.32 $ \\\hline
\end{tabular}
\caption{\label{tab:CuxBi2Se3_Rvalues_3axis}Refinements in space group $ R\overline{3}m $ and $ C2/m $ for different samples and temperatures. In order to compare the reliability values, the refinements in space group $ C2/m $ were carried out with the same structural parameters as in $ R\overline{3}m $ (left of the arrow "$ \rightarrow $") and then with the
structural parameters refined according to space group $ C2/m $ (right of the arrow "$ \rightarrow $"). The $ \mathrm{w}R(\mathrm{all}) $ values improve by approximately $ 0.5\, \% $; all reliability values are given in  \% .}
\end{table}

The anisotropic displacement parameters obtained in the refinements of the  $ R\overline{3}m $ and $ C2/m $ structures with the five data sets are shown in Table V. In both structures, we keep the symmetry constraints of the rhombohedral structure. The displacement parameters are quite anisotropic with larger $U_{33}$ values, which reflects the
layered character of the Bi$_2$Se$_3$ crystal structure. Strong bonds only exist within the quintuples layers, while the inter-layer bonding only arises from the
van der Waals potential. The refinements indicate soft $c$ polarized phonons.

\begin{table}
\begin{tabular}{c c c c c c c}\hline
\\[-1em]
                                             &                                       &                   & \multicolumn{2}{c}{$ R\overline{3}m $} & \multicolumn{2}{c}{$ C2/m $} \\
                                             &                                       &                   & $ U_{11} $  & $ U_{33} $ & $ U_{11} $ & $ U_{33} $  \\\hline
 \multirow{9}{*}{\rotatebox{90}{Sample: S1}} & \multirow{3}{*}{$ 1.9\,\mathrm{K} $}  & $ \mathrm{Bi1} $  & $ 42(4) $&  $ 119(7)  $  & $ 25(3) $&  $ 118(3) $ \\
                                             &                                       & $ \mathrm{Se2} $  & $ 45(5) $&  $ 100(10) $  & $ 27(5) $&  $ 112(5) $ \\
                                             &                                       & $ \mathrm{Se3} $  & $ 39(4) $&  $ 128(9)  $  & $ 43(4) $&  $ 113(4) $ \\\cline{2-7}
                                             & \multirow{3}{*}{$ 4.4\,\mathrm{K} $}  & $ \mathrm{Bi1} $  & $ 97(5) $&  $ 188(10) $  & $ 76(4) $&  $ 181(6) $ \\
                                             &                                       & $ \mathrm{Se2} $  & $ 95(6) $&  $ 188(13) $  & $ 73(6) $&  $ 192(9) $ \\
                                             &                                       & $ \mathrm{Se3} $  & $ 99(5) $&  $ 198(12) $  & $ 97(5) $&  $ 172(7) $ \\\cline{2-7}
                                             & \multirow{3}{*}{$ 300\,\mathrm{K} $}  & $ \mathrm{Bi1} $  & $ 155(7)$&  $ 334(13) $  & $ 141(5)$&  $ 331(6) $ \\
                                             &                                       & $ \mathrm{Se2} $  & $ 132(9)$&  $ 255(15) $  & $ 122(9)$&  $ 259(9) $ \\
                                             &                                       & $ \mathrm{Se3} $  & $ 147(8)$&  $ 287(13) $  & $ 147(8)$&  $ 277(7) $ \\\hline
                                             & \\\hline
 \multirow{6}{*}{\rotatebox{90}{Sample: S3}} & \multirow{3}{*}{$ 2\,\mathrm{K} $}    & $ \mathrm{Bi1} $  & $ 13(4) $& $ 100(6)  $  & $ 14(2) $& $ 91(2)   $  \\
                                             &                                       & $ \mathrm{Se2} $  & $ 10(4) $& $ 73(8)   $  & $ 16(3) $& $ 68(3)   $ \\
                                             &                                       & $ \mathrm{Se3} $  & $ 16(4) $& $ 103(7)  $  & $ 21(2) $& $ 95(2)   $   \\\cline{2-7}
                                             & \multirow{3}{*}{$ 300\,\mathrm{K} $}  & $ \mathrm{Bi1} $  & $ 83(7) $& $ 212(12) $  & $ 78(4) $& $ 197(5)  $  \\
                                             &                                       & $ \mathrm{Se2} $  & $ 68(9) $& $ 120(15) $  & $ 69(7) $& $ 117(7)  $  \\
                                             &                                       & $ \mathrm{Se3} $  & $ 77(8) $& $ 159(13) $  & $ 85(5) $& $ 145(5)  $  \\\hline
\end{tabular}
\caption{\label{anisoADP} Anisotropic displacement parameters of the atoms of Bi$_2$Se$_3$ of the samples S1 and S3 at different
temperatures refined in space groups $R\overline{3}m$ and $C2/m$. The constraints for ADPs in space group $R\overline{3}m$, $U_{11}=U_{22}=2U_{12}$ and $U_{13}=U_{23}=0$
are also used in $C2/m$, and all parameters are given in 10$^{-4}$\AA$^2$.}
\end{table}

\subsection{Analysis of the resemblance of Bragg scattering data}

We have analyzed the resemblance of Bragg scattering data taken as function of temperature by calculating the correlation function for data taken at temperatures $T_1$ and $T_2$ \cite{Bikondoa2017a}:

\begin{eqnarray}\label{eq:correlation_coefficient}
r(T_1, T_2) = \frac{\overline{I_{T_1}I_{T_2}} - \overline{I_{T_1}}~\overline{I_{T_2}}}{\sqrt{\left(\overline{I_{T_1}^2} - \overline{I_{T_1}}^2\right) \left(\overline{I_{T_2}^2} - \overline{I_{T_2}}^2 \right)}}
\end{eqnarray}

where $ I_T $ is the number of counts at temperature $ T $ and the vinculum denotes the mean with respect to all pixels and $ \omega $ positions.

The correlation function defined in equation \ref{eq:correlation_coefficient} does not depend on the total intensity of the signal, but only on the distribution along the $ \omega $ scan and the pixels. If the peak shape or position differ at two temperatures $ T_1 $ and $ T_2 $, the quantity $ r(T_1, T_2) $ is small.

In Fig. 6 we show colour maps of the $ r(T_1, T_2) $ values plotted against the two temperatures.
It can be seen that there is a strong tendency that the detector images are similar for similar temperatures and differ more strongly if the temperature difference is greater. If the detector images changed suddenly, there would be two temperature intervals, where correlation coefficients are high for temperatures that lie in the same interval and lower correlation coefficients for temperatures in different intervals.
We conclude that also the shape and position of the reflections show no indication for a structural phase transition.

\begin{figure}
\includegraphics[width=0.9\columnwidth]{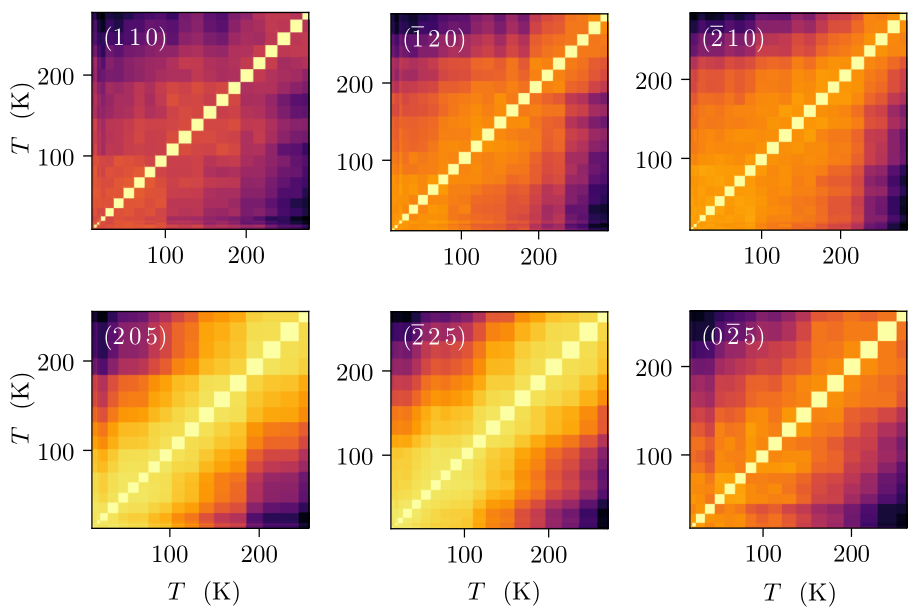}
  \caption{\label{fig:correlation}Color maps of the correlation function of the Bragg scattering data taken at two sets of equivalent reflections; the correlation function is plotted against the two temperatures to be compared. For all reflections there is no indication for a discontinuity of the correlation function as it is expected for a structural phase transition.
 }
  \label{disp}
 \end{figure}

\subsection{Calculated scattering density maps}

Extra scattering arising from the inserted atoms can be visualized in the scattering maps obtained by Fourier transformation.
Such calculated maps are shown in Fig. 7 for Bragg reflection data taken with sample S1 at 1.9\,K. However, there is no evidence
for extra scattering in these maps, which only show the expected peaks at the Bi and Se positions. Fourier maps calculated for the other sample S3 at $ 2\,\mathrm{K} $ also do not indicate extra scattering.

\begin{figure}
\includegraphics[width=0.9\columnwidth]{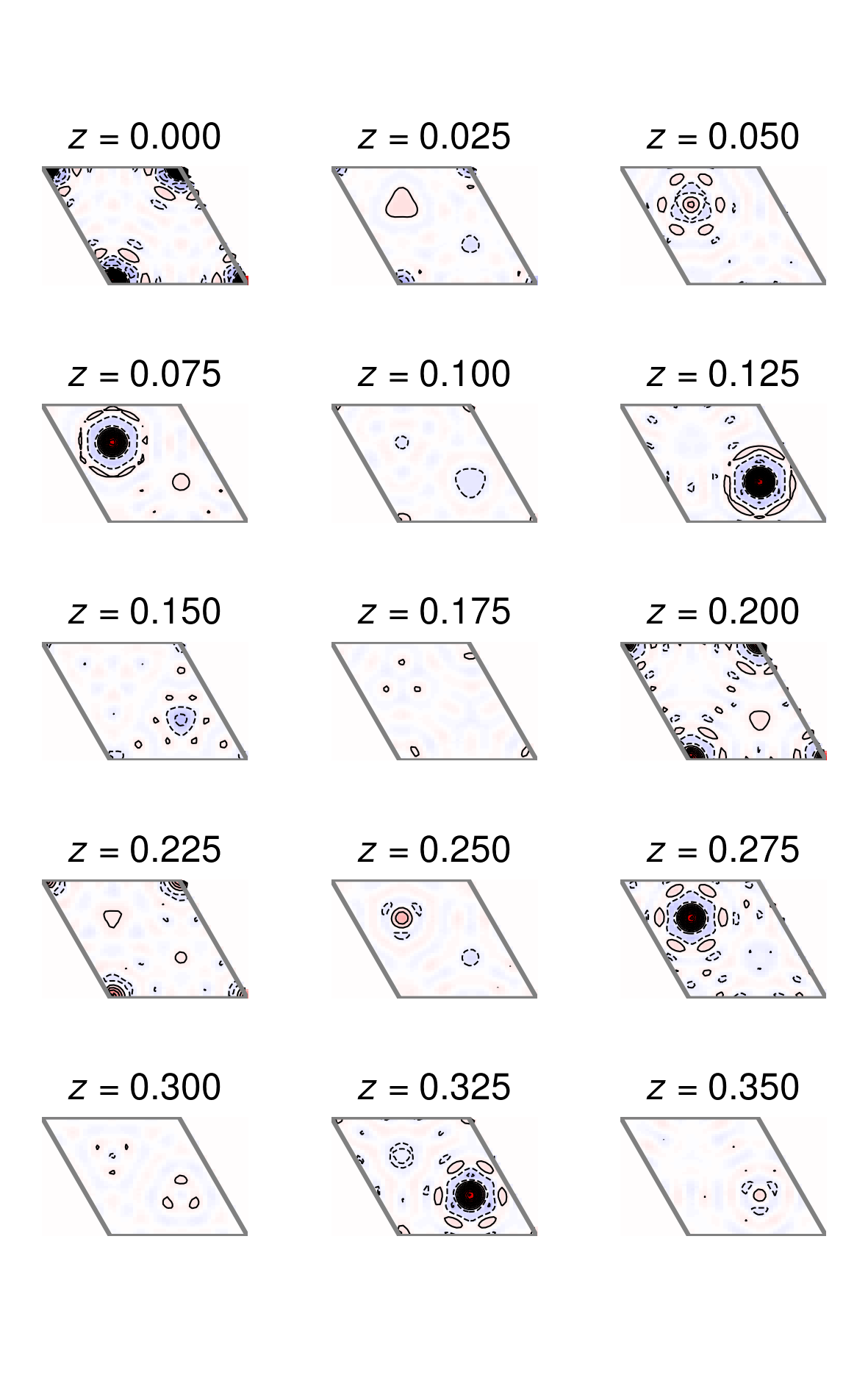}
  \caption{\label{structure}
Total scattering maps calculated with data set taken on sample S1 at $ 1.9\,\mathrm{K} $. The Bi and Se atoms are clearly visible, but there is no evidence for any
extra scattering in these maps.
 }
  \label{disp}
 \end{figure}

\bibliographystyle{apsrev4-1}
%


\begin{thebibliography}{32}%
\makeatletter
\providecommand \@ifxundefined [1]{%
 \@ifx{#1\undefined}
}%
\providecommand \@ifnum [1]{%
 \ifnum #1\expandafter \@firstoftwo
 \else \expandafter \@secondoftwo
 \fi
}%
\providecommand \@ifx [1]{%
 \ifx #1\expandafter \@firstoftwo
 \else \expandafter \@secondoftwo
 \fi
}%
\providecommand \natexlab [1]{#1}%
\providecommand \enquote  [1]{``#1''}%
\providecommand \bibnamefont  [1]{#1}%
\providecommand \bibfnamefont [1]{#1}%
\providecommand \citenamefont [1]{#1}%
\providecommand \href@noop [0]{\@secondoftwo}%
\providecommand \href [0]{\begingroup \@sanitize@url \@href}%
\providecommand \@href[1]{\@@startlink{#1}\@@href}%
\providecommand \@@href[1]{\endgroup#1\@@endlink}%
\providecommand \@sanitize@url [0]{\catcode `\\12\catcode `\$12\catcode
  `\&12\catcode `\#12\catcode `\^12\catcode `\_12\catcode `\%12\relax}%
\providecommand \@@startlink[1]{}%
\providecommand \@@endlink[0]{}%
\providecommand \url  [0]{\begingroup\@sanitize@url \@url }%
\providecommand \@url [1]{\endgroup\@href {#1}{\urlprefix }}%
\providecommand \urlprefix  [0]{URL }%
\providecommand \Eprint [0]{\href }%
\providecommand \doibase [0]{http://dx.doi.org/}%
\providecommand \selectlanguage [0]{\@gobble}%
\providecommand \bibinfo  [0]{\@secondoftwo}%
\providecommand \bibfield  [0]{\@secondoftwo}%
\providecommand \translation [1]{[#1]}%
\providecommand \BibitemOpen [0]{}%
\providecommand \bibitemStop [0]{}%
\providecommand \bibitemNoStop [0]{.\EOS\space}%
\providecommand \EOS [0]{\spacefactor3000\relax}%
\providecommand \BibitemShut  [1]{\csname bibitem#1\endcsname}%
\let\auto@bib@innerbib\@empty
\bibitem [{\citenamefont {Hor}\ \emph {et~al.}(2010)\citenamefont {Hor},
  \citenamefont {Williams}, \citenamefont {Checkelsky}, \citenamefont
  {Roushan}, \citenamefont {Seo}, \citenamefont {Xu}, \citenamefont
  {Zandbergen}, \citenamefont {Yazdani}, \citenamefont {Ong},\ and\
  \citenamefont {Cava}}]{Hor2010}%
  \BibitemOpen
  \bibfield  {author} {\bibinfo {author} {\bibfnamefont {Y.~S.}\ \bibnamefont
  {Hor}}, \bibinfo {author} {\bibfnamefont {A.~J.}\ \bibnamefont {Williams}},
  \bibinfo {author} {\bibfnamefont {J.~G.}\ \bibnamefont {Checkelsky}},
  \bibinfo {author} {\bibfnamefont {P.}~\bibnamefont {Roushan}}, \bibinfo
  {author} {\bibfnamefont {J.}~\bibnamefont {Seo}}, \bibinfo {author}
  {\bibfnamefont {Q.}~\bibnamefont {Xu}}, \bibinfo {author} {\bibfnamefont
  {H.~W.}\ \bibnamefont {Zandbergen}}, \bibinfo {author} {\bibfnamefont
  {A.}~\bibnamefont {Yazdani}}, \bibinfo {author} {\bibfnamefont {N.~P.}\
  \bibnamefont {Ong}}, \ and\ \bibinfo {author} {\bibfnamefont {R.~J.}\
  \bibnamefont {Cava}},\ }\href {\doibase 10.1103/PhysRevLett.104.057001}
  {\bibfield  {journal} {\bibinfo  {journal} {Phys. Rev. Lett.}\ }\textbf
  {\bibinfo {volume} {104}},\ \bibinfo {pages} {057001} (\bibinfo {year}
  {2010})}\BibitemShut {NoStop}%
\bibitem [{\citenamefont {Fu}\ and\ \citenamefont {Berg}(2010)}]{Fu2010}%
  \BibitemOpen
  \bibfield  {author} {\bibinfo {author} {\bibfnamefont {L.}~\bibnamefont
  {Fu}}\ and\ \bibinfo {author} {\bibfnamefont {E.}~\bibnamefont {Berg}},\
  }\href {\doibase 10.1103/PhysRevLett.105.097001} {\bibfield  {journal}
  {\bibinfo  {journal} {Phys. Rev. Lett.}\ }\textbf {\bibinfo {volume} {105}},\
  \bibinfo {pages} {097001} (\bibinfo {year} {2010})}\BibitemShut {NoStop}%
\bibitem [{\citenamefont {Liu}\ \emph {et~al.}(2015)\citenamefont {Liu},
  \citenamefont {Yao}, \citenamefont {Shao}, \citenamefont {Zuo}, \citenamefont
  {Pi}, \citenamefont {Tan}, \citenamefont {Zhang},\ and\ \citenamefont
  {Zhang}}]{Liu2015}%
  \BibitemOpen
  \bibfield  {author} {\bibinfo {author} {\bibfnamefont {Z.}~\bibnamefont
  {Liu}}, \bibinfo {author} {\bibfnamefont {X.}~\bibnamefont {Yao}}, \bibinfo
  {author} {\bibfnamefont {J.}~\bibnamefont {Shao}}, \bibinfo {author}
  {\bibfnamefont {M.}~\bibnamefont {Zuo}}, \bibinfo {author} {\bibfnamefont
  {L.}~\bibnamefont {Pi}}, \bibinfo {author} {\bibfnamefont {S.}~\bibnamefont
  {Tan}}, \bibinfo {author} {\bibfnamefont {C.}~\bibnamefont {Zhang}}, \ and\
  \bibinfo {author} {\bibfnamefont {Y.}~\bibnamefont {Zhang}},\ }\href@noop {}
  {\bibfield  {journal} {\bibinfo  {journal} {Journal of the Americal Chemical
  Society}\ }\textbf {\bibinfo {volume} {\textbf{137}}},\ \bibinfo {pages}
  {10512} (\bibinfo {year} {2015})}\BibitemShut {NoStop}%
\bibitem [{\citenamefont {Shruti}\ \emph {et~al.}(2015)\citenamefont {Shruti},
  \citenamefont {Maurya}, \citenamefont {Neha}, \citenamefont {Srivastava},\
  and\ \citenamefont {Patnaik}}]{Shruti2015}%
  \BibitemOpen
  \bibfield  {author} {\bibinfo {author} {\bibfnamefont {V.~K.}\ \bibnamefont
  {Shruti}}, \bibinfo {author} {\bibfnamefont {P.}~\bibnamefont {Maurya}},
  \bibinfo {author} {\bibfnamefont {P.}~\bibnamefont {Neha}}, \bibinfo {author}
  {\bibfnamefont {P.}~\bibnamefont {Srivastava}}, \ and\ \bibinfo {author}
  {\bibfnamefont {S.}~\bibnamefont {Patnaik}},\ }\href@noop {} {\bibfield
  {journal} {\bibinfo  {journal} {Physical Review B}\ }\textbf {\bibinfo
  {volume} {\textbf{92}}},\ \bibinfo {pages} {020506} (\bibinfo {year}
  {2015})}\BibitemShut {NoStop}%
\bibitem [{\citenamefont {Qiu}\ \emph {et~al.}(2015)\citenamefont {Qiu},
  \citenamefont {Sanders}, \citenamefont {Dai}, \citenamefont {Medvedeva},
  \citenamefont {Wu}, \citenamefont {Ghaemi}, \citenamefont {Vojta},\ and\
  \citenamefont {Hor}}]{Qiu2015}%
  \BibitemOpen
  \bibfield  {author} {\bibinfo {author} {\bibfnamefont {Y.}~\bibnamefont
  {Qiu}}, \bibinfo {author} {\bibfnamefont {K.~N.}\ \bibnamefont {Sanders}},
  \bibinfo {author} {\bibfnamefont {J.}~\bibnamefont {Dai}}, \bibinfo {author}
  {\bibfnamefont {J.~E.}\ \bibnamefont {Medvedeva}}, \bibinfo {author}
  {\bibfnamefont {W.}~\bibnamefont {Wu}}, \bibinfo {author} {\bibfnamefont
  {P.}~\bibnamefont {Ghaemi}}, \bibinfo {author} {\bibfnamefont
  {T.}~\bibnamefont {Vojta}}, \ and\ \bibinfo {author} {\bibfnamefont {Y.~S.}\
  \bibnamefont {Hor}},\ }\href@noop {} {\bibfield  {journal} {\bibinfo
  {journal} {arXiv 1512.03519}\ } (\bibinfo {year} {2015})}\BibitemShut
  {NoStop}%
\bibitem [{\citenamefont {Matano}\ \emph {et~al.}(2016)\citenamefont {Matano},
  \citenamefont {Kriener}, \citenamefont {Segawa}, \citenamefont {Ando},\ and\
  \citenamefont {Zheng}}]{matano2016}%
  \BibitemOpen
  \bibfield  {author} {\bibinfo {author} {\bibfnamefont {K.}~\bibnamefont
  {Matano}}, \bibinfo {author} {\bibfnamefont {M.}~\bibnamefont {Kriener}},
  \bibinfo {author} {\bibfnamefont {K.}~\bibnamefont {Segawa}}, \bibinfo
  {author} {\bibfnamefont {Y.}~\bibnamefont {Ando}}, \ and\ \bibinfo {author}
  {\bibfnamefont {G.}~\bibnamefont {Zheng}},\ }\href@noop {} {\bibfield
  {journal} {\bibinfo  {journal} {Nature Physics}\ }\textbf {\bibinfo {volume}
  {{\bf 12}}},\ \bibinfo {pages} {852} (\bibinfo {year} {2016})}\BibitemShut
  {NoStop}%
\bibitem [{\citenamefont {Yonezawa}(2018)}]{Yonezawa2018}%
  \BibitemOpen
  \bibfield  {author} {\bibinfo {author} {\bibfnamefont {S.}~\bibnamefont
  {Yonezawa}},\ }\href {\doibase 10.3390/condmat4010002} {\bibfield  {journal}
  {\bibinfo  {journal} {Condensed Matter}\ }\textbf {\bibinfo {volume} {4}},\
  \bibinfo {pages} {2} (\bibinfo {year} {2018})}\BibitemShut {NoStop}%
\bibitem [{\citenamefont {Yonezawa}\ \emph {et~al.}(2017)\citenamefont
  {Yonezawa}, \citenamefont {Tajiri}, \citenamefont {Nakata}, \citenamefont
  {Nagai}, \citenamefont {Wang}, \citenamefont {Segawa}, \citenamefont {Ando},\
  and\ \citenamefont {Maeno}}]{Yonezawa2017}%
  \BibitemOpen
  \bibfield  {author} {\bibinfo {author} {\bibfnamefont {S.}~\bibnamefont
  {Yonezawa}}, \bibinfo {author} {\bibfnamefont {K.}~\bibnamefont {Tajiri}},
  \bibinfo {author} {\bibfnamefont {S.}~\bibnamefont {Nakata}}, \bibinfo
  {author} {\bibfnamefont {Y.}~\bibnamefont {Nagai}}, \bibinfo {author}
  {\bibfnamefont {Z.}~\bibnamefont {Wang}}, \bibinfo {author} {\bibfnamefont
  {K.}~\bibnamefont {Segawa}}, \bibinfo {author} {\bibfnamefont
  {Y.}~\bibnamefont {Ando}}, \ and\ \bibinfo {author} {\bibfnamefont
  {Y.}~\bibnamefont {Maeno}},\ }\href {\doibase 10.1038/nphys3907} {\bibfield
  {journal} {\bibinfo  {journal} {Nature Physics}\ }\textbf {\bibinfo {volume}
  {13}},\ \bibinfo {pages} {123} (\bibinfo {year} {2017})}\BibitemShut
  {NoStop}%
\bibitem [{\citenamefont {Asaba}\ \emph {et~al.}(2017)\citenamefont {Asaba},
  \citenamefont {Lawson}, \citenamefont {Tinsman}, \citenamefont {Chen},
  \citenamefont {Corbae}, \citenamefont {Li}, \citenamefont {Qiu},
  \citenamefont {Hor}, \citenamefont {Fu},\ and\ \citenamefont
  {Li}}]{Asaba2017}%
  \BibitemOpen
  \bibfield  {author} {\bibinfo {author} {\bibfnamefont {T.}~\bibnamefont
  {Asaba}}, \bibinfo {author} {\bibfnamefont {B.~J.}\ \bibnamefont {Lawson}},
  \bibinfo {author} {\bibfnamefont {C.}~\bibnamefont {Tinsman}}, \bibinfo
  {author} {\bibfnamefont {L.}~\bibnamefont {Chen}}, \bibinfo {author}
  {\bibfnamefont {P.}~\bibnamefont {Corbae}}, \bibinfo {author} {\bibfnamefont
  {G.}~\bibnamefont {Li}}, \bibinfo {author} {\bibfnamefont {Y.}~\bibnamefont
  {Qiu}}, \bibinfo {author} {\bibfnamefont {Y.~S.}\ \bibnamefont {Hor}},
  \bibinfo {author} {\bibfnamefont {L.}~\bibnamefont {Fu}}, \ and\ \bibinfo
  {author} {\bibfnamefont {L.}~\bibnamefont {Li}},\ }\href@noop {} {\bibfield
  {journal} {\bibinfo  {journal} {Physical Review X}\ }\textbf {\bibinfo
  {volume} {\textbf{7}}},\ \bibinfo {pages} {011009} (\bibinfo {year}
  {2017})}\BibitemShut {NoStop}%
\bibitem [{\citenamefont {Pan}\ \emph {et~al.}(2016)\citenamefont {Pan},
  \citenamefont {Nikitin}, \citenamefont {Araizi}, \citenamefont {Huang},
  \citenamefont {Matsushita}, \citenamefont {Naka},\ and\ \citenamefont
  {de~Visser}}]{Pan2016}%
  \BibitemOpen
  \bibfield  {author} {\bibinfo {author} {\bibfnamefont {Y.}~\bibnamefont
  {Pan}}, \bibinfo {author} {\bibfnamefont {A.~M.}\ \bibnamefont {Nikitin}},
  \bibinfo {author} {\bibfnamefont {G.~K.}\ \bibnamefont {Araizi}}, \bibinfo
  {author} {\bibfnamefont {Y.~K.}\ \bibnamefont {Huang}}, \bibinfo {author}
  {\bibfnamefont {Y.}~\bibnamefont {Matsushita}}, \bibinfo {author}
  {\bibfnamefont {T.}~\bibnamefont {Naka}}, \ and\ \bibinfo {author}
  {\bibfnamefont {A.}~\bibnamefont {de~Visser}},\ }\href {\doibase
  10.1038/srep28632} {\bibfield  {journal} {\bibinfo  {journal} {Scientific
  Reports}\ }\textbf {\bibinfo {volume} {6}},\ \bibinfo {pages} {28632}
  (\bibinfo {year} {2016})}\BibitemShut {NoStop}%
\bibitem [{\citenamefont {Du}\ \emph {et~al.}(2017)\citenamefont {Du},
  \citenamefont {Li}, \citenamefont {Schneeloch}, \citenamefont {Zhong},
  \citenamefont {Gu}, \citenamefont {Yang}, \citenamefont {Lin},\ and\
  \citenamefont {Wen}}]{Du2017}%
  \BibitemOpen
  \bibfield  {author} {\bibinfo {author} {\bibfnamefont {G.}~\bibnamefont
  {Du}}, \bibinfo {author} {\bibfnamefont {Y.}~\bibnamefont {Li}}, \bibinfo
  {author} {\bibfnamefont {J.}~\bibnamefont {Schneeloch}}, \bibinfo {author}
  {\bibfnamefont {R.~D.}\ \bibnamefont {Zhong}}, \bibinfo {author}
  {\bibfnamefont {G.}~\bibnamefont {Gu}}, \bibinfo {author} {\bibfnamefont
  {H.}~\bibnamefont {Yang}}, \bibinfo {author} {\bibfnamefont {H.}~\bibnamefont
  {Lin}}, \ and\ \bibinfo {author} {\bibfnamefont {H.-H.}\ \bibnamefont
  {Wen}},\ }\href {\doibase 10.1007/s11433-016-0499-x} {\bibfield  {journal}
  {\bibinfo  {journal} {Science China Physics, Mechanics {\&} Astronomy}\
  }\textbf {\bibinfo {volume} {60}},\ \bibinfo {pages} {037411} (\bibinfo
  {year} {2017})}\BibitemShut {NoStop}%
\bibitem [{\citenamefont {Shen}\ \emph {et~al.}(2017)\citenamefont {Shen},
  \citenamefont {He}, \citenamefont {Yuan}, \citenamefont {Huang},
  \citenamefont {Cho}, \citenamefont {Lee}, \citenamefont {Hor}, \citenamefont
  {Law},\ and\ \citenamefont {Lortz}}]{Shen2017}%
  \BibitemOpen
  \bibfield  {author} {\bibinfo {author} {\bibfnamefont {J.}~\bibnamefont
  {Shen}}, \bibinfo {author} {\bibfnamefont {W.-Y.}\ \bibnamefont {He}},
  \bibinfo {author} {\bibfnamefont {N.~F.~Q.}\ \bibnamefont {Yuan}}, \bibinfo
  {author} {\bibfnamefont {Z.}~\bibnamefont {Huang}}, \bibinfo {author}
  {\bibfnamefont {C.-w.}\ \bibnamefont {Cho}}, \bibinfo {author} {\bibfnamefont
  {S.~H.}\ \bibnamefont {Lee}}, \bibinfo {author} {\bibfnamefont {Y.~S.}\
  \bibnamefont {Hor}}, \bibinfo {author} {\bibfnamefont {K.~T.}\ \bibnamefont
  {Law}}, \ and\ \bibinfo {author} {\bibfnamefont {R.}~\bibnamefont {Lortz}},\
  }\href {\doibase 10.1038/s41535-017-0064-1} {\bibfield  {journal} {\bibinfo
  {journal} {npj Quantum Materials}\ }\textbf {\bibinfo {volume} {2}},\
  \bibinfo {pages} {59} (\bibinfo {year} {2017})}\BibitemShut {NoStop}%
\bibitem [{\citenamefont {Smylie}\ \emph {et~al.}(2018)\citenamefont {Smylie},
  \citenamefont {Willa}, \citenamefont {Claus}, \citenamefont {Koshelev},
  \citenamefont {Song}, \citenamefont {Kwok}, \citenamefont {Islam},
  \citenamefont {Gu}, \citenamefont {Schneeloch}, \citenamefont {Zhong},\ and\
  \citenamefont {Welp}}]{Smylie2018}%
  \BibitemOpen
  \bibfield  {author} {\bibinfo {author} {\bibfnamefont {M.~P.}\ \bibnamefont
  {Smylie}}, \bibinfo {author} {\bibfnamefont {K.}~\bibnamefont {Willa}},
  \bibinfo {author} {\bibfnamefont {H.}~\bibnamefont {Claus}}, \bibinfo
  {author} {\bibfnamefont {A.~E.}\ \bibnamefont {Koshelev}}, \bibinfo {author}
  {\bibfnamefont {K.~W.}\ \bibnamefont {Song}}, \bibinfo {author}
  {\bibfnamefont {W.-K.}\ \bibnamefont {Kwok}}, \bibinfo {author}
  {\bibfnamefont {Z.}~\bibnamefont {Islam}}, \bibinfo {author} {\bibfnamefont
  {G.~D.}\ \bibnamefont {Gu}}, \bibinfo {author} {\bibfnamefont {J.~A.}\
  \bibnamefont {Schneeloch}}, \bibinfo {author} {\bibnamefont {Zhong}}, \ and\
  \bibinfo {author} {\bibfnamefont {U.}~\bibnamefont {Welp}},\ }\href@noop {}
  {\bibfield  {journal} {\bibinfo  {journal} {Scientific Reports}\ }\textbf
  {\bibinfo {volume} {\textbf{8}}},\ \bibinfo {pages} {7666} (\bibinfo {year}
  {2018})}\BibitemShut {NoStop}%
\bibitem [{\citenamefont {Wang}\ \emph {et~al.}(2011)\citenamefont {Wang},
  \citenamefont {Y.}, \citenamefont {Jiang}, \citenamefont {Liu}, \citenamefont
  {Chang}, \citenamefont {Chen}, \citenamefont {Li}, \citenamefont {Song},
  \citenamefont {Wang}, \citenamefont {He}, \citenamefont {Chen}, \citenamefont
  {Duan}, \citenamefont {Xue},\ and\ \citenamefont {Ma}}]{Wang2011}%
  \BibitemOpen
  \bibfield  {author} {\bibinfo {author} {\bibfnamefont {Y.-L.}\ \bibnamefont
  {Wang}}, \bibinfo {author} {\bibfnamefont {X.}~\bibnamefont {Y.}}, \bibinfo
  {author} {\bibfnamefont {Y.-P.}\ \bibnamefont {Jiang}}, \bibinfo {author}
  {\bibfnamefont {J.-W.}\ \bibnamefont {Liu}}, \bibinfo {author} {\bibfnamefont
  {C.-Z.}\ \bibnamefont {Chang}}, \bibinfo {author} {\bibfnamefont
  {M.}~\bibnamefont {Chen}}, \bibinfo {author} {\bibfnamefont {Z.}~\bibnamefont
  {Li}}, \bibinfo {author} {\bibfnamefont {C.-L.}\ \bibnamefont {Song}},
  \bibinfo {author} {\bibfnamefont {L.-L.}\ \bibnamefont {Wang}}, \bibinfo
  {author} {\bibfnamefont {K.}~\bibnamefont {He}}, \bibinfo {author}
  {\bibfnamefont {X.}~\bibnamefont {Chen}}, \bibinfo {author} {\bibfnamefont
  {W.-H.}\ \bibnamefont {Duan}}, \bibinfo {author} {\bibfnamefont {Q.-K.}\
  \bibnamefont {Xue}}, \ and\ \bibinfo {author} {\bibfnamefont {X.-C.}\
  \bibnamefont {Ma}},\ }\href@noop {} {\bibfield  {journal} {\bibinfo
  {journal} {Physical Review B}\ }\textbf {\bibinfo {volume} {\textbf{84}}},\
  \bibinfo {pages} {075335} (\bibinfo {year} {2011})}\BibitemShut {NoStop}%
\bibitem [{\citenamefont {Kriener}\ \emph
  {et~al.}(2011{\natexlab{a}})\citenamefont {Kriener}, \citenamefont {Segawa},
  \citenamefont {Ren}, \citenamefont {Sasaki},\ and\ \citenamefont
  {Ando}}]{Kriener2011}%
  \BibitemOpen
  \bibfield  {author} {\bibinfo {author} {\bibfnamefont {M.}~\bibnamefont
  {Kriener}}, \bibinfo {author} {\bibfnamefont {K.}~\bibnamefont {Segawa}},
  \bibinfo {author} {\bibfnamefont {Z.}~\bibnamefont {Ren}}, \bibinfo {author}
  {\bibfnamefont {S.}~\bibnamefont {Sasaki}}, \ and\ \bibinfo {author}
  {\bibfnamefont {Y.}~\bibnamefont {Ando}},\ }\href {\doibase
  10.1103/PhysRevLett.106.127004} {\bibfield  {journal} {\bibinfo  {journal}
  {Phys. Rev. Lett.}\ }\textbf {\bibinfo {volume} {106}},\ \bibinfo {pages}
  {127004} (\bibinfo {year} {2011}{\natexlab{a}})}\BibitemShut {NoStop}%
\bibitem [{\citenamefont {Kriener}\ \emph
  {et~al.}(2011{\natexlab{b}})\citenamefont {Kriener}, \citenamefont {Segawa},
  \citenamefont {Ren}, \citenamefont {Sasaki}, \citenamefont {Wada},
  \citenamefont {Kuwabata},\ and\ \citenamefont {Ando}}]{Kriener2011a}%
  \BibitemOpen
  \bibfield  {author} {\bibinfo {author} {\bibfnamefont {M.}~\bibnamefont
  {Kriener}}, \bibinfo {author} {\bibfnamefont {K.}~\bibnamefont {Segawa}},
  \bibinfo {author} {\bibfnamefont {Z.}~\bibnamefont {Ren}}, \bibinfo {author}
  {\bibfnamefont {S.}~\bibnamefont {Sasaki}}, \bibinfo {author} {\bibfnamefont
  {S.}~\bibnamefont {Wada}}, \bibinfo {author} {\bibfnamefont {S.}~\bibnamefont
  {Kuwabata}}, \ and\ \bibinfo {author} {\bibfnamefont {Y.}~\bibnamefont
  {Ando}},\ }\href {\doibase 10.1103/PhysRevB.84.054513} {\bibfield  {journal}
  {\bibinfo  {journal} {Phys. Rev. B}\ }\textbf {\bibinfo {volume} {84}},\
  \bibinfo {pages} {054513} (\bibinfo {year} {2011}{\natexlab{b}})}\BibitemShut
  {NoStop}%
\bibitem [{\citenamefont {Nakajima}(1963)}]{Nakajima1963}%
  \BibitemOpen
  \bibfield  {author} {\bibinfo {author} {\bibfnamefont {S.}~\bibnamefont
  {Nakajima}},\ }\href@noop {} {\bibfield  {journal} {\bibinfo  {journal} {J.
  Phys. Chem. Solids}\ }\textbf {\bibinfo {volume} {\textbf{24}}},\ \bibinfo
  {pages} {479} (\bibinfo {year} {1963})}\BibitemShut {NoStop}%
\bibitem [{\citenamefont {Li}\ \emph {et~al.}(2018)\citenamefont {Li},
  \citenamefont {Wang}, \citenamefont {Zhang}, \citenamefont {Feng},
  \citenamefont {Jiang}, \citenamefont {Han}, \citenamefont {Chen},
  \citenamefont {Ye}, \citenamefont {Gao}, \citenamefont {Jia}, \citenamefont
  {Li}, \citenamefont {Qiao}, \citenamefont {Qian}, \citenamefont {Xu},
  \citenamefont {Tian},\ and\ \citenamefont {Gao}}]{Li2018}%
  \BibitemOpen
  \bibfield  {author} {\bibinfo {author} {\bibfnamefont {Z.}~\bibnamefont
  {Li}}, \bibinfo {author} {\bibfnamefont {M.}~\bibnamefont {Wang}}, \bibinfo
  {author} {\bibfnamefont {D.}~\bibnamefont {Zhang}}, \bibinfo {author}
  {\bibfnamefont {N.}~\bibnamefont {Feng}}, \bibinfo {author} {\bibfnamefont
  {W.}~\bibnamefont {Jiang}}, \bibinfo {author} {\bibfnamefont
  {C.}~\bibnamefont {Han}}, \bibinfo {author} {\bibfnamefont {W.}~\bibnamefont
  {Chen}}, \bibinfo {author} {\bibfnamefont {M.}~\bibnamefont {Ye}}, \bibinfo
  {author} {\bibfnamefont {C.}~\bibnamefont {Gao}}, \bibinfo {author}
  {\bibfnamefont {J.}~\bibnamefont {Jia}}, \bibinfo {author} {\bibfnamefont
  {J.}~\bibnamefont {Li}}, \bibinfo {author} {\bibfnamefont {S.}~\bibnamefont
  {Qiao}}, \bibinfo {author} {\bibfnamefont {D.}~\bibnamefont {Qian}}, \bibinfo
  {author} {\bibfnamefont {B.}~\bibnamefont {Xu}}, \bibinfo {author}
  {\bibfnamefont {H.}~\bibnamefont {Tian}}, \ and\ \bibinfo {author}
  {\bibfnamefont {B.}~\bibnamefont {Gao}},\ }\href@noop {} {\bibfield
  {journal} {\bibinfo  {journal} {Physcal Review Materials}\ }\textbf {\bibinfo
  {volume} {\textbf{2}}},\ \bibinfo {pages} {014201} (\bibinfo {year}
  {2018})}\BibitemShut {NoStop}%
\bibitem [{\citenamefont {Kuntsevich}\ \emph {et~al.}(2018)\citenamefont
  {Kuntsevich}, \citenamefont {Bryzgalov}, \citenamefont {Prudkoglyad},
  \citenamefont {Martovitskii}, \citenamefont {Selivanov},\ and\ \citenamefont
  {Chizhevskii}}]{Kuntsevich2018}%
  \BibitemOpen
  \bibfield  {author} {\bibinfo {author} {\bibfnamefont {A.~Y.}\ \bibnamefont
  {Kuntsevich}}, \bibinfo {author} {\bibfnamefont {M.~A.}\ \bibnamefont
  {Bryzgalov}}, \bibinfo {author} {\bibfnamefont {V.~A.}\ \bibnamefont
  {Prudkoglyad}}, \bibinfo {author} {\bibfnamefont {V.~P.}\ \bibnamefont
  {Martovitskii}}, \bibinfo {author} {\bibfnamefont {Y.~G.}\ \bibnamefont
  {Selivanov}}, \ and\ \bibinfo {author} {\bibfnamefont {E.~G.}\ \bibnamefont
  {Chizhevskii}},\ }\href@noop {} {\bibfield  {journal} {\bibinfo  {journal}
  {New Journal of Physics}\ }\textbf {\bibinfo {volume} {\textbf{20}}},\
  \bibinfo {pages} {103022} (\bibinfo {year} {2018})}\BibitemShut {NoStop}%
\bibitem [{d9g()}]{d9guide}%
  \BibitemOpen
  \href
  {https://www.ill.eu/users/instruments/instruments-list/d9/documentation/}
  { {\bibinfo {note} {D9: Hot Neutron Four-Circle Diffractometer}}},\
  \bibinfo {organization} {Institut Laue-Langevin},\ \bibinfo {address} {71
  avenue des Martyrs CS 20156, 38042 Grenoble Cedex 9, France}\BibitemShut
  {NoStop}%
\bibitem [{dat()}]{data}%
  \BibitemOpen
  \href@noop {} {}\bibinfo {note} {Data is available at
  https://doi.ill.fr/10.5291/ILL-DATA.5-41-914 and
  https://doi.ill.fr/10.5291/ILL-DATA.5-11-426}\BibitemShut {NoStop}%
\bibitem [{\citenamefont {Albinati}\ and\ \citenamefont {{\it et
  al.}}(2004)}]{tablesC}%
  \BibitemOpen
  \bibfield  {author} {\bibinfo {author} {\bibfnamefont {A.}~\bibnamefont
  {Albinati {\it et al.}}},\
  }\href@noop {} { {\bibinfo {note} {International Tables for
  Crystallography}}},\ edited by\ \bibinfo {editor} {\bibfnamefont
  {E.}~\bibnamefont {Prince}},\ Vol.~\bibinfo {volume} {C}\ (\bibinfo
  {publisher} {Kluwer Academic Publishers},\ \bibinfo {year}
  {2004})\BibitemShut {NoStop}%
\bibitem [{\citenamefont {P\'erez~Vicente}\ \emph {et~al.}(1999)\citenamefont
  {P\'erez~Vicente}, \citenamefont {Tirado}, \citenamefont {Adouby},
  \citenamefont {Jumas}, \citenamefont {Tour\'e},\ and\ \citenamefont
  {Kra}}]{Vicente1999}%
  \BibitemOpen
  \bibfield  {author} {\bibinfo {author} {\bibfnamefont {C.}~\bibnamefont
  {P\'erez~Vicente}}, \bibinfo {author} {\bibfnamefont {J.~L.}\ \bibnamefont
  {Tirado}}, \bibinfo {author} {\bibfnamefont {K.}~\bibnamefont {Adouby}},
  \bibinfo {author} {\bibfnamefont {J.~C.}\ \bibnamefont {Jumas}}, \bibinfo
  {author} {\bibfnamefont {A.~A.}\ \bibnamefont {Tour\'e}}, \ and\ \bibinfo
  {author} {\bibfnamefont {G.}~\bibnamefont {Kra}},\ }\href {\doibase
  10.1021/ic9812858} {\bibfield  {journal} {\bibinfo  {journal} {Inorganic
  Chemistry}\ }\textbf {\bibinfo {volume} {38}},\ \bibinfo {pages} {2131}
  (\bibinfo {year} {1999})} \BibitemShut {NoStop}%
\bibitem [{\citenamefont {Sobczak}\ \emph {et~al.}(2018)\citenamefont
  {Sobczak}, \citenamefont {Strak}, \citenamefont {Kempisty}, \citenamefont
  {Wolos}, \citenamefont {Hruban}, \citenamefont {Materna},\ and\ \citenamefont
  {Borysiuk}}]{Sobczak2018a}%
  \BibitemOpen
  \bibfield  {author} {\bibinfo {author} {\bibfnamefont {K.}~\bibnamefont
  {Sobczak}}, \bibinfo {author} {\bibfnamefont {P.}~\bibnamefont {Strak}},
  \bibinfo {author} {\bibfnamefont {P.}~\bibnamefont {Kempisty}}, \bibinfo
  {author} {\bibfnamefont {A.}~\bibnamefont {Wolos}}, \bibinfo {author}
  {\bibfnamefont {A.}~\bibnamefont {Hruban}}, \bibinfo {author} {\bibfnamefont
  {A.}~\bibnamefont {Materna}}, \ and\ \bibinfo {author} {\bibfnamefont
  {J.}~\bibnamefont {Borysiuk}},\ }\href@noop {} {\bibfield  {journal}
  {\bibinfo  {journal} {Physical Review Materials}\ }\textbf {\bibinfo {volume}
  {\textbf{2}}},\ \bibinfo {pages} {044203} (\bibinfo {year}
  {2018})}\BibitemShut {NoStop}%
\bibitem [{\citenamefont {Pet\v{r}\'{i}\v{c}ek}\ \emph
  {et~al.}(2014)\citenamefont {Pet\v{r}\'{i}\v{c}ek}, \citenamefont
  {Du\v{s}ek},\ and\ \citenamefont {Palatinus}}]{petricek}%
  \BibitemOpen
  \bibfield  {author} {\bibinfo {author} {\bibfnamefont {V.}~\bibnamefont
  {Pet\v{r}\'{i}\v{c}ek}}, \bibinfo {author} {\bibfnamefont {M.}~\bibnamefont
  {Du\v{s}ek}}, \ and\ \bibinfo {author} {\bibfnamefont {L.}~\bibnamefont
  {Palatinus}},\ }\href@noop {} {\bibfield  {journal} {\bibinfo  {journal} {Z.
  Kristallogr.}\ }\textbf {\bibinfo {volume} {229(5)}},\ \bibinfo {pages} {345}
  (\bibinfo {year} {2014})}\BibitemShut {NoStop}%
\bibitem{note-calc}{In order to calculate the mean Cu occupation we add the contributions
from the Cu sites (i) to (iv) and average over the five data sets.}
\bibitem [{\citenamefont {Arnold}\ and\ \citenamefont {{\it et
  al.}}(2005)}]{tablesA}%
  \BibitemOpen
  \bibfield  {author} {\bibinfo {author} {\bibfnamefont {H.}~\bibnamefont
  {Arnold{\it et al.}}},\ }\href@noop
  {} { {\bibinfo {note} {International Tables for Crystallography}}},\
  edited by\ \bibinfo {editor} {\bibfnamefont {T.}~\bibnamefont {Hahn}},\
  Vol.~\bibinfo {volume} {A}\ (\bibinfo  {publisher} {Springer},\ \bibinfo
  {year} {2005})\BibitemShut {NoStop}%
\bibitem{Willa2018}{K. Willa, R. Willa, Kok Wee Song, G. D. Gu, J. A. Schneeloch, R. Zhong, A. E. Koshelev, Wai-Kwong Kwok, and U. Welp,
Phys. Rev. B {\bf 98}, 184509 (2018).}  
\bibitem [{\citenamefont {Stokes}\ and\ \citenamefont {Hatch}(1988)}]{stokes}%
  \BibitemOpen
  \bibfield  {author} {\bibinfo {author} {\bibfnamefont {H.~T.}\ \bibnamefont
  {Stokes}}\ and\ \bibinfo {author} {\bibfnamefont {D.~M.}\ \bibnamefont
  {Hatch}},\ }\href@noop {} { {\bibinfo {note} {Isotropy Subgroups of the
  230 Crystallographic Space Groups}}}\ (\bibinfo  {publisher} {World
  Scientific, Singapore},\ \bibinfo {year} {1988})\BibitemShut {NoStop}%
\bibitem [{\citenamefont {Andersen}\ \emph {et~al.}(2018)\citenamefont
  {Andersen}, \citenamefont {Wang}, \citenamefont {Lorenz},\ and\ \citenamefont
  {Ando}}]{Andersen2018}%
  \BibitemOpen
  \bibfield  {author} {\bibinfo {author} {\bibfnamefont {L.}~\bibnamefont
  {Andersen}}, \bibinfo {author} {\bibfnamefont {Z.}~\bibnamefont {Wang}},
  \bibinfo {author} {\bibfnamefont {T.}~\bibnamefont {Lorenz}}, \ and\ \bibinfo
  {author} {\bibfnamefont {Y.}~\bibnamefont {Ando}},\ }\href {\doibase
  10.1103/PhysRevB.98.220512} {\bibfield  {journal} {\bibinfo  {journal} {Phys.
  Rev. B}\ }\textbf {\bibinfo {volume} {98}},\ \bibinfo {pages} {220512}
  (\bibinfo {year} {2018})}\BibitemShut {NoStop}%
\bibitem{Venderbos2016}
{J. W. F. Venderbos, V. Kozii, and Liang Fu, Phys. Rev. B {\bf 94}, 094522 (2016).}
\bibitem{Cho2019}
{Chang-woo Cho, Junying Shen, Jian Lyu, S. H. Lee, Yew San Hor, M. Hecker, J. Schmalian, and R. Lortz, arXiv:1905.01702.}
\bibitem [{org()}]{orgel1959}%
  \BibitemOpen
  \href@noop {} {}\bibinfo {note} {L. E. Orgel, J. Chem. Soc. 3815
  (1959).}\BibitemShut {Stop}%
\bibitem [{gal()}]{galy1975}%
  \BibitemOpen
  \href@noop {} {}\bibinfo {note} {J. Galy, and G. Meunier, J. of Sol. State
  Chemistry {\bf 13}, 142 (1975).}\BibitemShut {Stop}%
\bibitem [{laa()}]{laarif1986}%
  \BibitemOpen
  \href@noop {} {}\bibinfo {note} {A. Laarif, and F. Theobald, Solid State
  Ionics {\bf 21}, 183 (1986).}\BibitemShut {Stop}%
\bibitem [{\citenamefont {Bikondoa}(2017)}]{Bikondoa2017a}%
  \BibitemOpen
  \bibfield  {author} {\bibinfo {author} {\bibfnamefont {O.}~\bibnamefont
  {Bikondoa}},\ }\href@noop {} {\bibfield  {journal} {\bibinfo  {journal}
  {Journal of Applied Crystallography}\ }\textbf {\bibinfo {volume}
  {\textbf{50}}},\ \bibinfo {pages} {357} (\bibinfo {year} {2017})}\BibitemShut
  {NoStop}%
\end{thebibliography}



\end{document}